\documentclass[a4paper,10pt]{article}

\usepackage{graphicx,amssymb,color}
\usepackage{epsfig,subfigure}

\graphicspath{{pictures/}}

\title{Modelling Erythroblastic Islands: Using a Hybrid Model to Assess the Function of Central Macrophage} 

\author{S. Fischer$^{1, \dagger}$, P. Kurbatova$^{2,3,\dagger}$, N. Bessonov$^{4}$, O. Gandrillon$^{3,5}$, \\ V. Volpert$^{2,3}$, F. Crauste$^{2,3,*}$}

\date{\today}

\begin{document}

\maketitle

\centerline{$^*$ Corresponding author}
\centerline{Tel: +33 472 448 516; Fax: +33 472 431 687; Email: crauste@math.univ-lyon1.fr}\vspace{1cm}

\centerline{$^1$ Universit\'e de Lyon, CNRS, INRIA, INSA-Lyon, LIRIS Combining, UMR5205}
\centerline{F-69621, France}\vspace{1ex}

\centerline{$^2$ Universit\'{e} de Lyon, Universit\'{e} Lyon 1, CNRS UMR 5208, Institut Camille Jordan}
\centerline{43 blvd du 11 novembre 1918, F-69622 Villeurbanne-Cedex, France}\vspace{1ex}

\centerline{$^3$ INRIA Team Dracula, INRIA Grenoble Rh\^one-Alpes Center}\vspace{1ex}

\centerline{$^4$ Institute of Mechanical Engineering Problems, 199178 Saint Petersburg, Russia}\vspace{1ex}

\centerline{$^5$ Universit\'{e} de Lyon, Universit\'{e} Lyon 1, CNRS UMR 5534, Centre de G\'{e}n\'{e}tique et de }
\centerline{Physiologie Mol\'{e}culaire et Cellulaire, F-69622
Villeurbanne-Cedex, France}\vspace{1cm}

\centerline{$^\dagger$ Both authors contributed equally to this work}\vspace{1cm}

\centerline{Short title: Modelling Erythroblastic Islands}\vspace{2cm}

\noindent \textit{Funding Statement:} This work has been partially supported by the ANR grant ANR-09-JCJC-0100-01, project ProCell, and the ANR Project Bimod. The funders had no role in study design, data collection and analysis, decision to publish, or preparation of the manuscript.\vspace{2cm}

\noindent \textit{Author Contributions:} Conceived and designed the experiments: SF, PK, VV. Performed the experiments: SF, PK. Analyzed the data: SF, PK, OG, VV, FC. Conceived and designed the mathematical models: SF, PK, OG, VV, FC. Designed and implemented algorithms: SF, PK, NB. Wrote the paper: SF, PK, OG, VV, FC.

\newpage

\begin{abstract}
The production and regulation of red blood cells, erythropoiesis, occurs in the bone marrow where erythroid cells proliferate and differentiate within particular structures, called erythroblastic islands. A typical structure of these islands consists in a macrophage (white cell) surrounded by immature erythroid cells (progenitors), with more mature cells on the periphery of the island, ready to leave the bone marrow and enter the bloodstream. A hybrid model, coupling a continuous model (ordinary differential equations) describing intracellular regulation through competition of two key proteins, to a discrete spatial model describing cell-cell interactions, with growth factor diffusion in the medium described by a continuous model (partial differential equations), is proposed to investigate the role of the central macrophage in normal erythropoiesis. Intracellular competition of the two proteins leads the erythroid cell to either proliferation, differentiation, or death by apoptosis. This approach allows considering spatial aspects of erythropoiesis, involved for instance in the occurrence of cellular interactions or the access to external factors, as well as dynamics of intracellular and extracellular scales of this complex cellular process, accounting for stochasticity in cell cycle durations and orientation of the mitotic spindle. The analysis of the model shows a strong effect of the central macrophage on the stability of an erythroblastic island, when assuming the macrophage releases pro-survival cytokines. Even though it is not clear whether or not erythroblastic island stability must be required, investigation of the model concludes that stability improves responsiveness of the model, hence stressing out the potential relevance of the central macrophage in normal erythropoiesis.
\end{abstract}

\newpage

\section*{Author Summary}

Red blood cells are produced in the bone marrow and released into the bloodstream. Different cell populations involved in their production (from the very immature ones to differentiated ones) are usually kept in a steady state through various and complex negative feedback controls. Inside the bone marrow, immature red blood cells mature and differentiate within particular structures, called erythroblastic islands, consisting of groups of cells surrounding a macrophage (a big white cell). The role of the macrophage as well as the different feedback mechanisms involved in red blood cell production (erythropoiesis) are not completely understood. We propose a new model of red blood cell production considering both mechanisms at the intracellular scale (protein regulation, cell fate decision) and at the cell population scale (cell-cell interactions, influence of feedback controls), and allowing the numerical simulation of an erythroblastic island. Our results bring new information on the role of the macrophage and the erythroblastic island in the production and regulation of red blood cells, the macrophage being shown to play a key role in the stability and robustness of the production process, and will improve our understanding of in vivo erythropoiesis.


\section{Introduction}

Hematopoiesis, the process of production and regulation of all blood cell types, has been the topic of modelling works for decades. Dynamics of hematopoietic stem cells have been described by Mackey's early works \cite{m1978,m1979}, and later developed by Mackey and co-authors \cite{mr1994, mr1999} as well as Loeffler and co-workers \cite{roeder2006, wl1985, wls1985}. The white blood cell production process (leukopoiesis) has been modelled in order to illustrate some pathological cases, such as effects of radiations \cite{wl1985, wls1985}, and bring some information on hematological diseases, mainly leukemia \cite{bbm2004, cm2005a, michor2009, michor2005, michor2007, michor2008, ScholzEngelLoeffler2005} and  cyclical neutropenia \cite{bbm2003, cm2005b, hdm1998}. The platelet production process (thrombopoiesis) attracted less attention through years \cite{eller-etal1987, wichmann-etal1979}, with however some modelling effort in the 2000's \cite{am2008, santillan-etal2000}. The red blood cell production process (erythropoiesis) has recently been the focus of modelling in hematopoiesis. Early works by Wichmann et al. \cite{wlpw1989}, Wulff et al. \cite{wwlp1989}, B\'{e}lair et al. \cite{bmm1995}, Mahaffy et al. \cite{mbm1998}, on the regulation of cell population kinetics (mainly erythroid progenitors and mature red blood cells) have been completed by recent works \cite{adit2006, acr2006b, bcst2004}. Particular attention was paid to the relevance of erythroid progenitor self-renewal in response to stress and considering multiscale approaches \cite{bcfkv, Bessonov, cpgmg2008, cdgv2010}, that is including both cell population kinetics and intracellular regulatory networks dynamics in the models, in order to give insight in the mechanisms involved in erythropoiesis. A case study of erythropoiesis is anemia \cite{savill-etal2009}, which allows observing the system in a stress situation that can be easily induced (at least in mice). A model of all hematopoietic cell lineages has been proposed by Colijn and Mackey \cite{cm2005a, cm2005b}, including dynamics of hematopoietic stem cells and white cell, red blood cell and platelet lineages.

Almost all above-mentioned contributions to hematopoiesis modelling are based on theoretical studies of deterministic models, and are mainly qualitative rather than quantitative. Other approaches, developed by Veng-Pedersen and co-workers \cite{chapel-etal2000, freise-etal2008, vengpedersen-etal2002, woo-etal2006, woo-etal2008}, focused on pharmacodynamics/pharmacokinetics (PK/PD) models related to erythropoiesis modelling. Such approaches are usually centered on parameter estimation and model fitting to data (statistical approaches). Nevertheless, none of the previously mentioned approaches did consider spatial aspects of hematopoiesis. Models describe cell population kinetics, either in the bone marrow or the spleen (where cell production and maturation occurs) or in the bloodstream (where differentiated and mature cells ultimately end up). Consequently, cellular regulation by cell-cell interaction was neither considered in these models.

We propose a new multiscale model for cell proliferation \cite{bcfkv, Bessonov}, applied to erythropoiesis in the bone marrow, based on hybrid modelling. This approach, taking into account both interactions at the cell population level and regulation at the intracellular level, allows studying cell proliferation at different scales. Moreover, the `hybrid' modelling consists in considering a continuous model at the intracellular scale (that is, deterministic or stochastic differential equations), where protein competition occurs, and a discrete model to describe cell evolution (every cell is a single object evolving off-lattice and interacting with surrounding cells and the medium), hence allowing considering small population effects as well as random effects. Hybrid multiscale models have been recently developed and used mainly, but not exclusively, to model solid tumor growth \cite{alarcon2009, hd2010, jeon-quaranta-cummings2010, Osborne, Patel, rccad2009, salazarciudad, spencer2006}.

Erythropoiesis consists in the differentiation and maturation of very immature blood cells, called progenitors, produced by differentiation of hematopoietic stem cells in the bone marrow. These erythroid progenitors differentiate through successive divisions and under the action of external signals into more and more mature cells, up to a stage called reticulocytes, which are almost differentiated red blood cells. After nuclei ejection, reticulocytes enter the bloodstream and finish their differentiation process to become erythrocytes (mature red blood cells). At every step of this differentiation process, erythroid cells can die by apoptosis (programmed cell death), and erythroid progenitors have been shown to be able to self-renew under stress conditions \cite{bauer1999, g2002, gsbs1999, pain1991}.  Moreover, cell fate is partially controlled by external feedback controls. For instance, death by apoptosis has been proved to be mainly negatively regulated by erythropoietin (Epo), a growth factor released by the kidneys when the organism lacks red blood cells \cite{kb1990}. Self-renewal is induced by glucocorticoids \cite{bauer1999, g2002, gsbs1999, pain1991}, but also by Epo \cite{Rubiolo, spivak1991} and some intracellular autocrine loops \cite{gsbs1999, sjh2000}.

In the hybrid model, cells are individual objects supposed to be able to either self-renew, differentiate or die by apoptosis, according to the state of the environment and the cell itself. Indeed, global feedback control mediated by the population of mature red blood cells (through erythropoietin release) is supposed to influence erythroid progenitor proliferation \cite{Rubiolo, spivak1991} and inhibit their death by apoptosis \cite{kb1990}. Local feedback control, mediated by reticulocytes and based upon Fas-ligand activity, is supposed to induce both differentiation and death by apoptosis \cite{maria}. Global and local feedback controls modify the activity of intracellular proteins. A set of two proteins, previously considered by the authors \cite{cdgv2010}, Erk and Fas, has been shown \cite{Rubiolo} to be involved in an antagonist loop where Erk, from the MAPK family, inhibits Fas (a TNF family member) and self-activates, high levels of Erk inducing cell proliferation depending on Epo levels, whereas Fas inhibits Erk and induces apoptosis and differentiation. Competition between these two proteins sets a relevant frame to observe, within a single cell, all three possible cell fates depending on the level of each protein.

In addition to intracellular regulation of cell fate and feedback induced either by mature red blood cells or reticulocytes on erythroid progenitors, an important aspect of erythropoiesis lays in the structure of the bone marrow. Erythroid progenitors are indeed produced in specific micro-environments, called niches \cite{Tsiftsoglou}, where erythroblastic islands develop. An erythroblastic island consists in a central macrophage surrounded by erythroid progenitors \cite{chasis2008}. The macrophage appears to act on surrounding cells, by affecting their proliferation and differentiation programs. For years, however, erythropoiesis has been studied under the influence of erythropoietin, which may induce differentiation and proliferation in vitro without the presence of the macrophage. Hence, the roles of the macrophage and the erythroblastic island have been more or less neglected. Consequently, spatial aspects of erythropoiesis (and hematopoiesis, in general) have usually not been considered when modelling cell population evolution or hematological diseases appearance and treatment.

We use a hybrid model to take into account both intracellular and extracellular regulation of erythropoiesis, but also mainly to study the importance of spatial structure of erythroblastic islands in the regulation of erythropoiesis. To our knowledge, this is the first attempt to model erythropoiesis by taking these aspects into account. We focus in particular on the role of the macrophage, by analyzing different situations, with and without macrophage at the center of the island, and actions of the feedback controls, especially regarding the feedback of the macrophage on erythroid cell proliferation and differentiation. The `classical' structure of the erythroblastic island, with the macrophage at the center and immature erythroid cells surrounding it, will be shown to make the hybrid model stable (so the island can be sustained for an unlimited number of cell cycles) and robust to perturbations due to global and local feedback controls. 

The next section is devoted to the presentation of the multiscale hybrid model. First, we focus on the intracellular scale, detailing regulatory networks considered in this work, described by nonlinear ordinary differential equations, and investigating their properties, in particular regarding the existence and stability of steady states associated to cell self-renewal, differentiation and death by apoptosis. Second, we concentrate on the extracellular scale, which consists in erythroid cell populations, possibly a macrophage, bone marrow medium, and is described by a computational model. Interactions between cells and with the surrounding medium are presented. Finally, we justify the coupling of both scales, through global and local feedback controls, and we discuss parameters of the model. In Section \ref{s-results} we present the analysis of the hybrid model. This analysis consists of two parts: first, we consider an erythroblastic island without a macrophage in its center, and investigate the stability of the island and the roles of feedback controls in the stability. Next we study an island consisting of immature erythroid cells surrounding a macrophage, and the roles of feedback controls in its stability. This analysis concludes to the central role of the macrophage in the stability of the erythroblastic island, and consequently of the relevance of considering spatial aspects when modeling erythropoiesis. 


\section{Model}

We introduce, in this section, the hybrid model for erythropoiesis used later for in silico experiments. It consists in the coupling of two models, with different space and time scales. The first model describes intracellular dynamics, represented by regulatory networks based on specific protein competition. This intracellular dynamical level can be easily modeled with a continuous approach, namely ordinary differential equations (ODEs) and/or partial differential equations (PDEs). The second model is at cell population level, where discrete cells and events are computed, so that stochasticity due to random events (cell cycle duration, orientation of the mitotic spindle at division) and small population effects plays an important role. Extracellular regulation by growth factors is partially due to continuous models (PDEs). Such discrete-continuous models are usually called hybrid models \cite{Bessonov, hd2010, Osborne, Patel, rccad2009, salazarciudad, spencer2006}. We hereafter present how the model is defined at each scale and how the different levels interact.

\subsection{Intracellular Scale: Mathematical Model}\label{mathematical}

The intracellular scale describes how each erythroid progenitor cell chooses between self-renewal, differentiation and
apoptosis. This choice depends on the level of expression of some proteins, involved in a regulatory network.

Ordinary differential equations are used to describe a simplified regulatory network, based on two competing proteins, Erk and Fas. Concentrations of the proteins evolve according to protein-related mechanisms as well as extracellular factor concentrations. A brief analysis of the dynamics of the system (\ref{erk})--(\ref{fas}) is performed in order to identify key factors in cell fate choice.


\subsubsection{ODE System}

Precise intracellular regulatory mechanisms involved in erythroid progenitor cell fate are largely unknown. Based on the current knowledge, we decided to focus on a simplified regulatory network based on two proteins, Erk and Fas, responsible respectively for cell self-renewal and proliferation and cell differentiation and apoptosis \cite{cpgmg2008, demin, maria, Rubiolo}. These proteins are antagonists: they inhibit each other's expression. They are also subject to external regulation, through feedback loops.

One of these feedback loops is based upon the population of reticulocytes (differentiated erythroid cells). They produce Fas-ligand which is fixed to their exterior cell membrane. Fas-ligand activates Fas, a transmembrane protein, and influences progenitor differentiation and apoptosis. Another feedback control is related to erythrocytes in bloodstream. Their quantity determines the release of erythropoietin and other hormones, called growth factors. Erythropoietin is known to inhibit erythroid progenitor apoptosis \cite{kb1990} and to stimulate immature erythroid progenitor self-renewal \cite{Rubiolo, spivak1991}. Other hormones, like glucocorticoids \cite{bauer1999, g2002, gsbs1999, pain1991}, also increase erythroid progenitor self-renewal by activating Erk.

The simplified system of Erk-Fas interactions considered as the main regulatory network for erythroid progenitor fate is then \cite{cdgv2010}
\begin{eqnarray}
\displaystyle\frac{dE}{dt} &=& (\alpha(Epo,GF) + \beta E^k) (1-E) - a E - b EF , \label{erk}\vspace{1ex}\\
\displaystyle\frac{dF}{dt} &=& \gamma(F_L) (1-F) - c EF - dF, \label{fas}
\end{eqnarray}
where $E$ and $F$ denote intracellular normalized levels of Erk and Fas. Equation (\ref{erk}) describes how Erk level evolves toward maximal value
1 by activation through hormones (function $\alpha$ of erythropoeitin, denoted by $Epo$, and other growth factors, denoted by $GF$) and self-activation
(parameters $\beta$ and $k$). In the meantime, Erk is linearly degraded with a rate $a$ and is inhibited by Fas with a rate $bF$.

Equation (\ref{fas}) is very similar, only there is no proof for Fas self-activation. Fas is however activitated by Fas-ligand, denoted by $F_L$, through the function $\gamma(F_L)$, it is degraded with a rate $d$ and inhibited by Erk with a rate $cE$.

This model simply describes a competition between two
proteins (Erk and Fas) through cross-inhibition, self-activation (only
for Erk), and activation by extracellular proteins.

In order to explore the parameter space efficiently and observe the possible
behaviors of the cell, it is relevant to study the mathematical
interpretation of parameters.

\subsubsection{Brief Analysis: Existence and Stability of Steady States}

For fixed values of $Epo$, $GF$ and $F_L$, (\ref{erk})-(\ref{fas}) is
a closed system of ordinary differential equations. In order to analyze the impact
of each parameter on the system's behavior, we can derive analytical forms of the zero-lines associated
with each equation. The zero-line $f$ associated with equation (\ref{erk}) is
\begin{equation}
\label{f}
f(E) =  \frac{1}{b}\left(\frac{\alpha}{E} - \beta E^{k-1} (1-E) - (a+\alpha)\right).
\end{equation}
The zero-line $g$ associated with equation (\ref{fas}) is
\begin{equation}
\label{g}
g(E) =  \displaystyle\frac{\displaystyle\frac{\gamma}{c}}{\displaystyle\frac{\gamma + d}{c} + E}.
\end{equation}

The shape of the zero-lines depends only on 4 and 2 parameters respectively ($\alpha/b$, $\beta/b$, $k$ and $a/b$ for $f$, $\gamma/c$ and $d/c$ for $g$), the 2
remaining parameters ($b$ and $c$, related to cross-inhibition) only affecting the absolute and relative speeds of the dynamics along each dimension.
The existence of steady states, defined as intersections of the two zero-lines, thus only depends on
the former 6 parameters.

Let us stress out that $\alpha$ represents the influence of external factors (erythropoietin, growth factors) on Erk activation, $\beta$ is related to Erk self-activation, and $\gamma$ describes the influence of Fas-ligand on Fas activation.

\begin{figure}[!ht]
 \centerline{\includegraphics[scale=0.45]{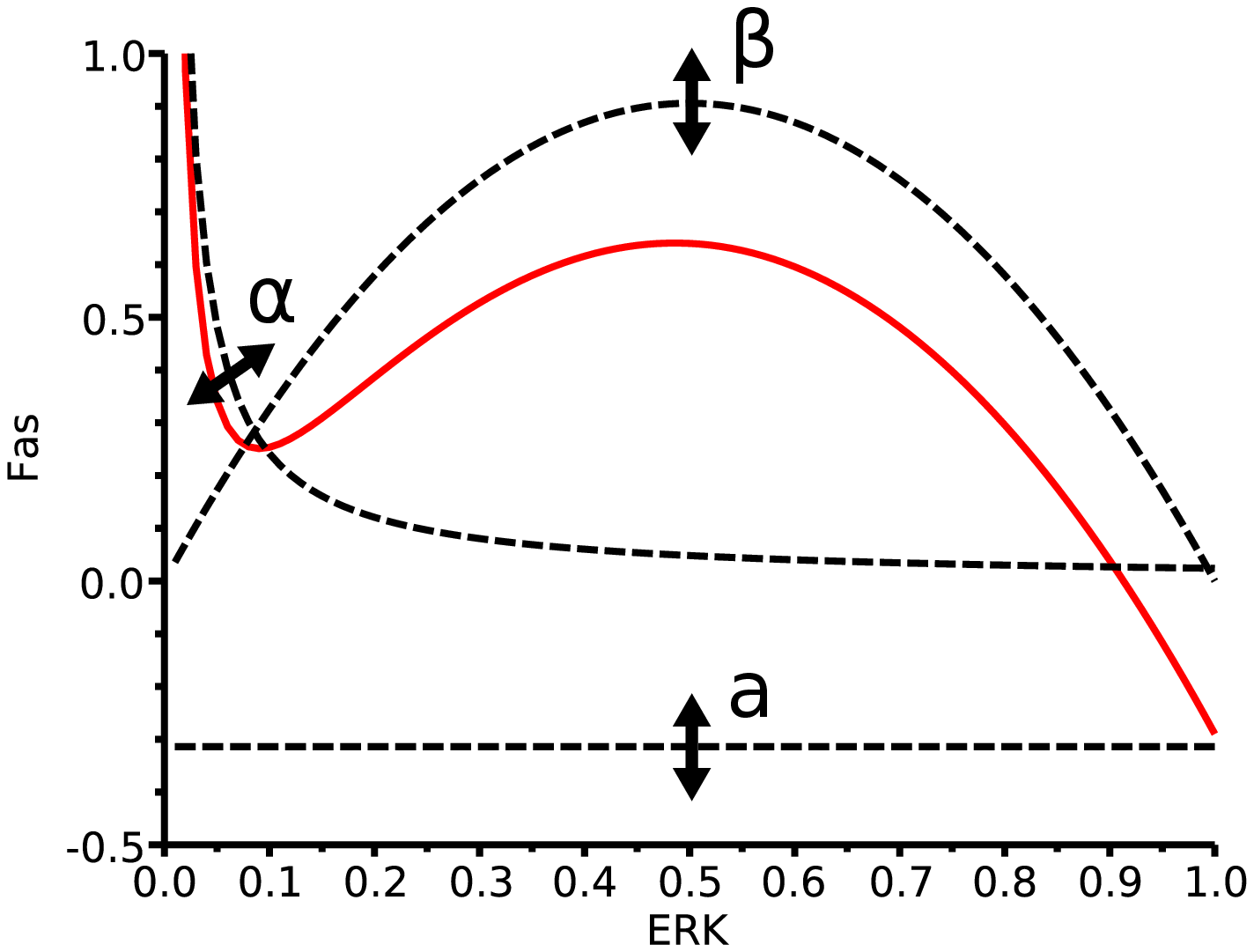}
 \includegraphics[scale=0.45]{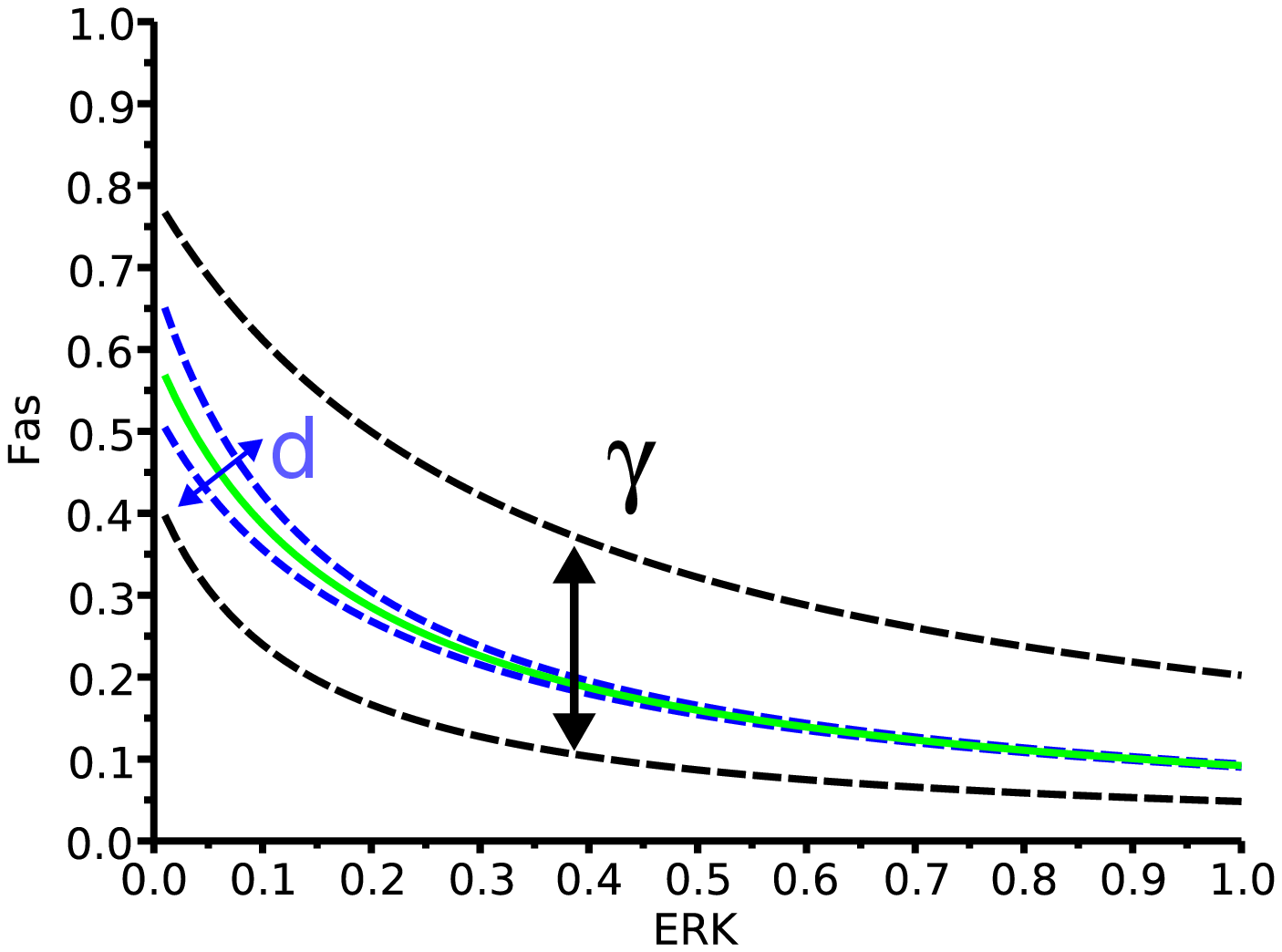}}
 \caption{\textbf{Zero-lines $f$ and $g$ defined in (\ref{f}) and (\ref{g}).} Left panel: the $f$ zero-line (straight red line) and the impact of parameters of equation (\ref{erk}). Right panel: the $g$ zero-line (straight green line) and the impact of parameters of equation (\ref{fas}).}
 \label{params}
\end{figure}

By considering that $f$ can be separated in 3 components, the
impact of each parameter can easily be observed. The same can be done for $g$. Figure \ref{params} shows that
the variations of the zero-lines are quite constrained, with a central role
of $\alpha$ and $\beta$ for $f$ and $\gamma$ for $g$. This will be studied in further detail in the following.

One may note that the effect of $k$ is not shown here, since $k=2$ in all experiments. It is however noticeable that increasing $k$
simply decreases the maximum value of the polynomial part of $f$ toward 0 while increasing the value for which it is reached toward 1.

Simple considerations show that the zero-lines can only cross in the
$[0,1]\times[0,1]$ domain. Furthermore, they intersect at least once, so there is at least one steady state, and
there can be only up to three steady states. When there is only one steady state, it is
stable. When there are three steady states, the central point is unstable and the two surrounding points are
stable \cite{cdgv2010}. The case with two steady states is singular so it is not really relevant.

\begin{figure}[!ht]
\begin{center}
\includegraphics[scale=0.43]{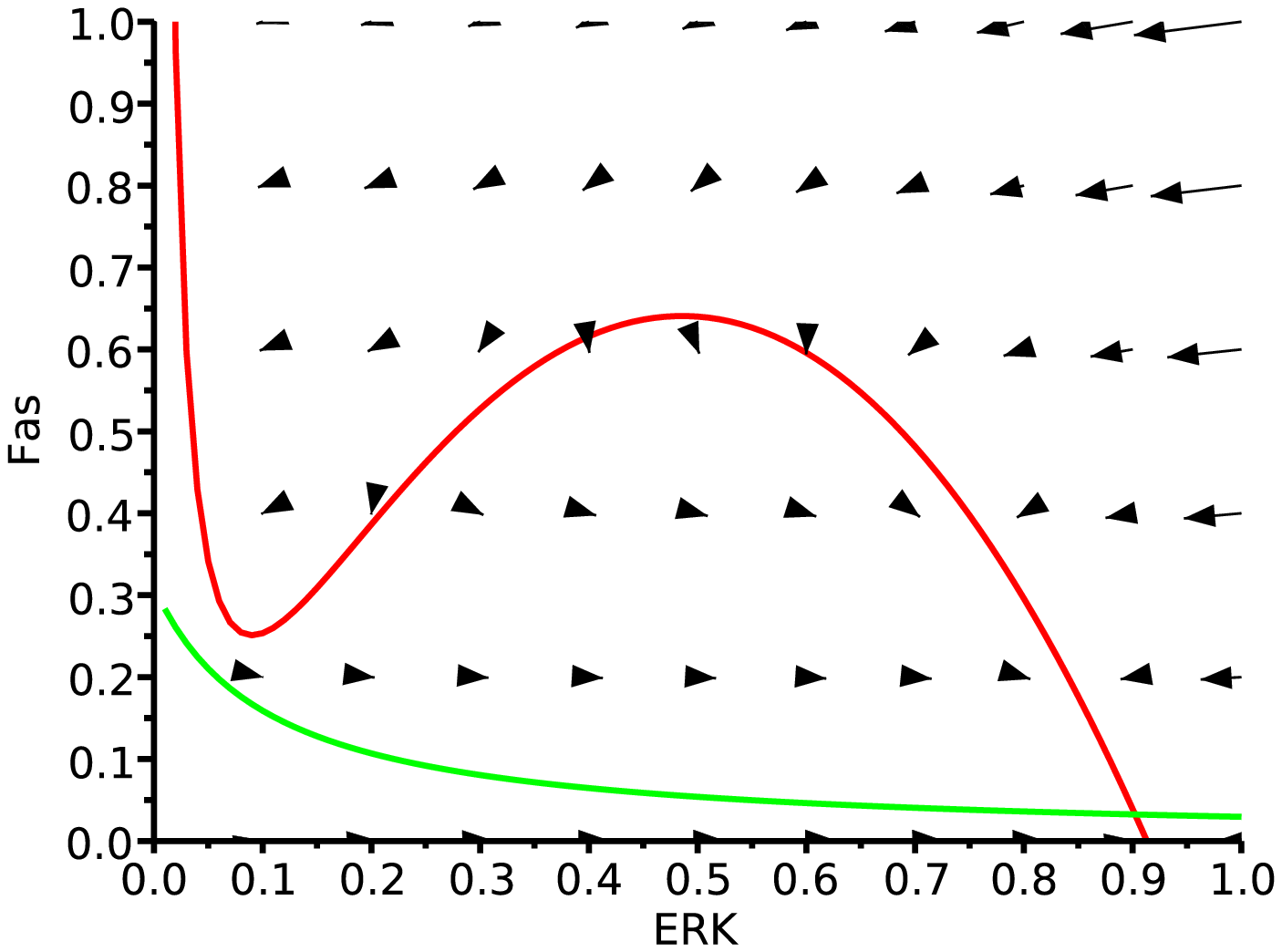}
\includegraphics[scale=0.43]{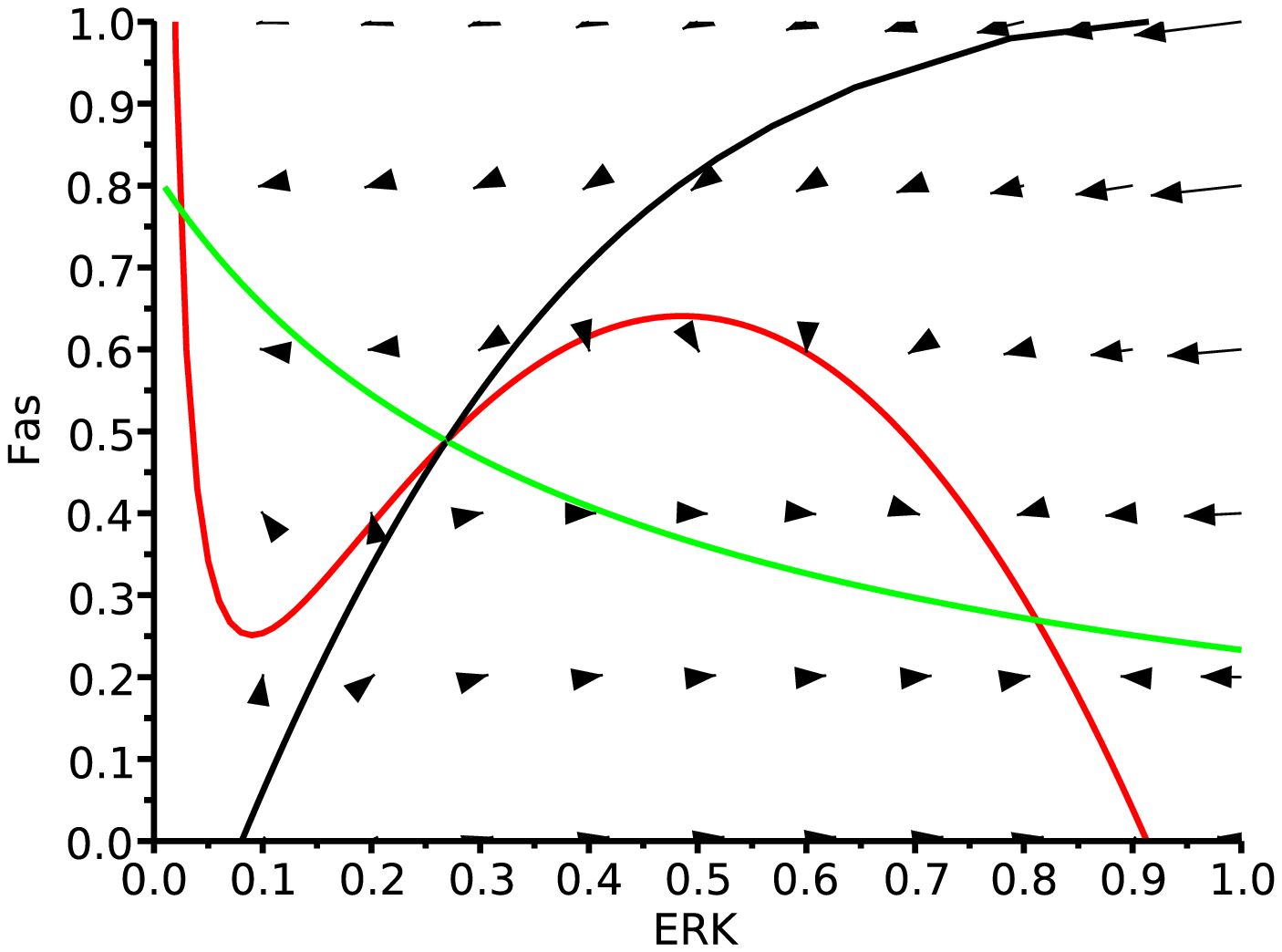}
\end{center}
A\hspace{6.7cm} B
\begin{center}
\includegraphics[scale=0.43]{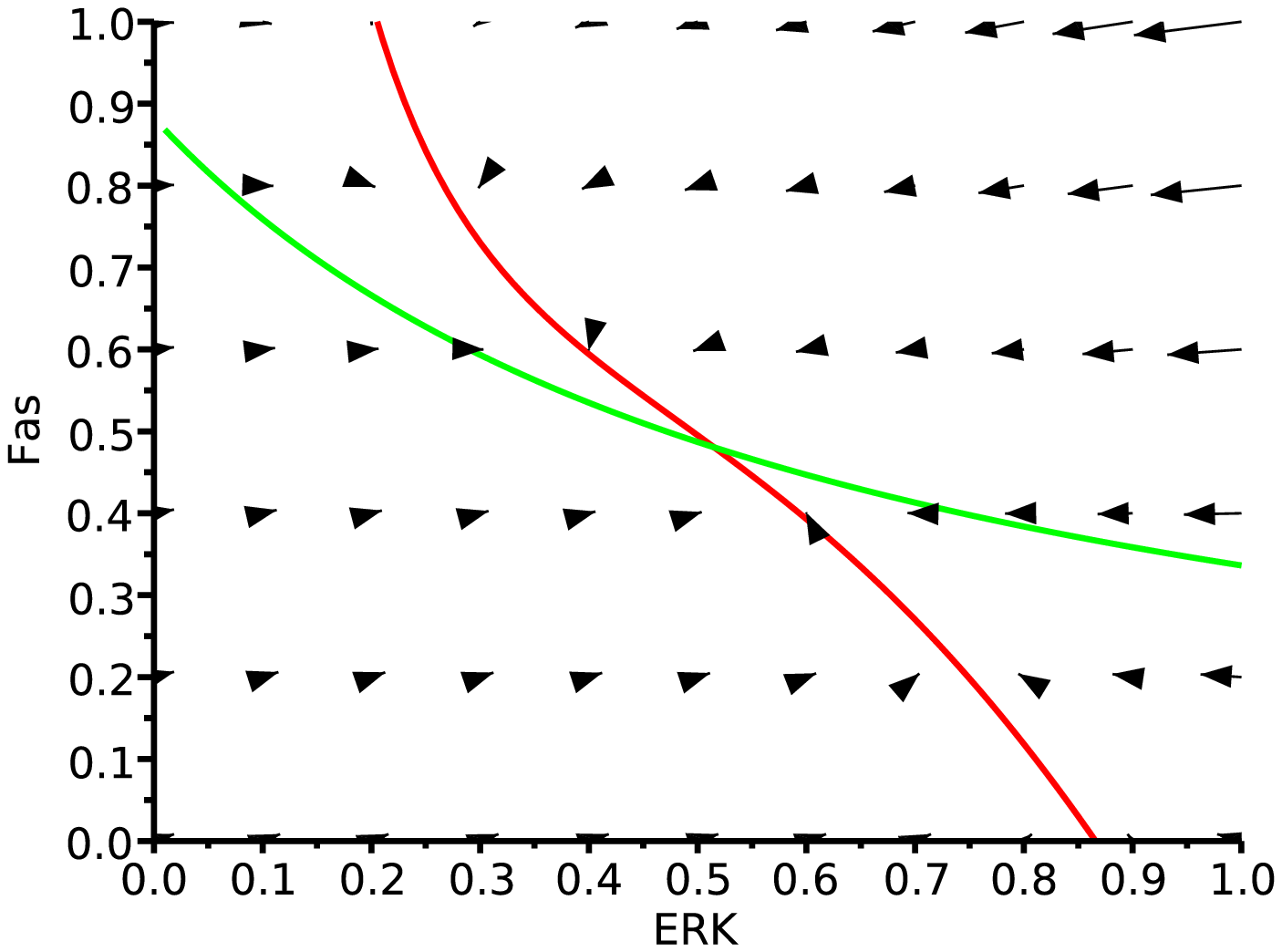}
\end{center}
\hspace{3.2cm} C
\caption{\textbf{Three possible and characteristic configurations of the system (\ref{f})--(\ref{g}).} The red line represents the $f$ zero-line, the green line the $g$ zero-line. Arrows indicate directions of the solutions of system (\ref{erk})--(\ref{fas}) in the (Erk,Fas) phase plane. A: Low value of $\alpha$, low value of $\gamma$. B: Low value of $\alpha$, medium value of $\gamma$ (the black curve is the separatrix). C: High value of $\alpha$ compared to $\beta$.}
 \label{bifurcations}
\end{figure}

Hence, the system exists in two configurations that we will call monostable (one stable
state) and bistable (two stable states). The next step is to determine how the system can go from one state to the other one. Thanks to Figure \ref{bifurcations}, two kind of bifurcations can be imagined to achieve this goal.

Let first start with the case where $\beta$ is large compared to $\alpha$, this means Erk self-activation rate is stronger than external activators (Epo for instance). In this case, $f$
has two local extrema (Figure \ref{bifurcations}.A). If $\gamma$, which is associated with Fas-ligand, varies then a double saddle-point bifurcation (or fold bifurcation)
occurs. For low and high values of Fas-ligand (that is low and high values of $\gamma$, respectively), there is only one stable steady state, whereas
for medium values of Fas ligand two stable points coexist, a separatrix determining
the attractive domain of each point (Figure \ref{bifurcations}.B).

Now consider the case where $\alpha$ is large compared to $\beta$. Then $f$ becomes strictly
decreasing and, if $\alpha$ is sufficiently large, there will be only one steady state
for every value of $\gamma$. The system is thus strictly monostable (Figure \ref{bifurcations}.C).

Finally, the system (\ref{f})--(\ref{g}) exists in two different cases: when $\alpha$ is low it is
locally bistable, depending on the value of $\gamma$, and when $\alpha$ is high,
it is strictly monostable. A kind of fork bifurcation occurs when $\alpha$ increases
(approximately around $\alpha=\beta / 27$, but exact value depends on other parameters
and is quite difficult to obtain).

\subsubsection{Dynamics of the ODE System}

Let us focus on the possible dynamics of the system (\ref{erk})--(\ref{fas}). This system must be able to divide the erythroid progenitor cell population into 3 different types: self-renewing, apoptotic and differentiating cells. Differentiation corresponds to low values of Fas and Erk at the end of the cycle, apoptosis to high values of Fas and self-renewal to high values of Erk and reasonable values of Fas.

A simple idea to achieve such a behavior is to introduce two threshold values. The first one, $F_{cr}$, induces apoptosis when reached. The second one, $E_{cr}$, determines the choice between self-renewal and differentiation for cells reaching the end of their cycle. So the phase plane of system (\ref{erk})--(\ref{fas}) can be divided in three parts: cells having low Erk and Fas values differentiate at the end of their cycle, cells having Erk values greater than $E_{cr}$ self-renew at the end of their cycle and cells reaching Fas values greater than $F_{cr}$ die immediately by apoptosis.

In order to get three robust cell subpopulations, it is necessary to determine how to correctly place these thresholds. Therefore, cases where the system is either strictly monostable or locally bistable have to be considered.

Let us recall that the position of the $f$ zero-line strongly depends on parameters $\alpha$ and $\beta$, whereas the position of the $g$ zero-line strongly depends on the parameter $\gamma$. Since $\beta$ denotes the self-activation rate of Erk and $\alpha$ represents a feedback control by global factors (this will be developed later in Section \ref{s-results}), then all erythroid progenitor cells will obey the same evolution rule for Erk. On the contrary, $\gamma$ represents a feedback control by a local factor (Fas-ligand emitted by reticulocytes and acting at short-range, see Section \ref{computational}) so all cell positions on the phase-plane will depend on their own exposition to Fas-ligand.

If the system is strictly monostable, then, independently from its starting position in the $(E,F)$-plane, a cell will converge toward a unique stable state, located on the $f$ zero-line and defined only by its exposition to Fas-ligand (the value of $\gamma$). So, at steady-state, if cells are exposed to a continuum of Fas-ligand values, they will be continuously located along the $f$ zero-line, and their level of Erk will not change: there is no real subpopulation. It is thus quite inefficient to work at steady-state (lack of robustness) to get three subpopulations corresponding to the three possible cell fates.

On the contrary, if the system is locally bistable, cells will still be asymptotically located along the $f$ zero-line, yet a part of it is unstable (roughly speaking the ascending part, see Figure \ref{params}) so cells will be separated into two subpopulations. In this case, the starting point is important: depending on where it is compared to the separatrix in the bistable domain, a cell will converge either toward the upper or the lower stable branch of $f$.

Based on these observations, we considered two different initial conditions and dynamics to get the three subpopulations.

\begin{figure}[!ht]
 \centerline{\includegraphics[scale=0.7]{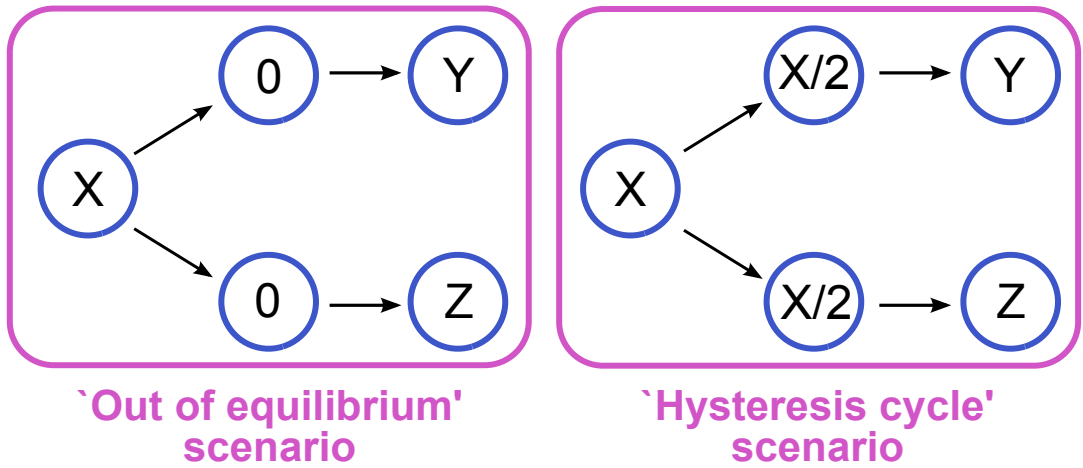}}
 \caption{\textbf{Schematic representation of the two scenarii considered for distribution of Erk and Fas quantities at cell division.} The left panel illustrates the `out of equilibrium' case, the right panel the `hysteresis cycle' case. In the `out of equilibrium' scenario a cell with a level $X=(E,F)$ of Erk and Fas at mitosis gives birth to two daughter cells with no Erk and no Fas ($(E,F)=(0,0)$). These cells will in turn reach different levels of Erk and Fas at the end of their own cell cycles (provided that they do not die), denoted by $Y$ and $Z$. In the `hysteresis cycle' scenario, a cell with a level $X=(E,F)$ of Erk and Fas produces two daughter cells with levels $X/2 = (E/2,F/2)$ which will reach a priori different levels of Erk and Fas at the end of their own cell cycle, due to stochasticity and response to signaling.}
 \label{fig-two-scenarii}
\end{figure}

The first idea is that, if working at steady-state does not allow to obtain distinct subpopulations, then working out of steady-state may be relevant. Consequently, in a first scenario called the `out of equilibrium case', all cells are arbitrarily put at the origin (that is $E=0$ and $F=0$) at the beginning of their cycle (see Figure \ref{fig-two-scenarii}, left) and, before they have reached steady-state, they are forced to take a decision based on Erk and Fas levels.

\begin{figure}[!ht]
\begin{center}
\includegraphics[scale=0.43]{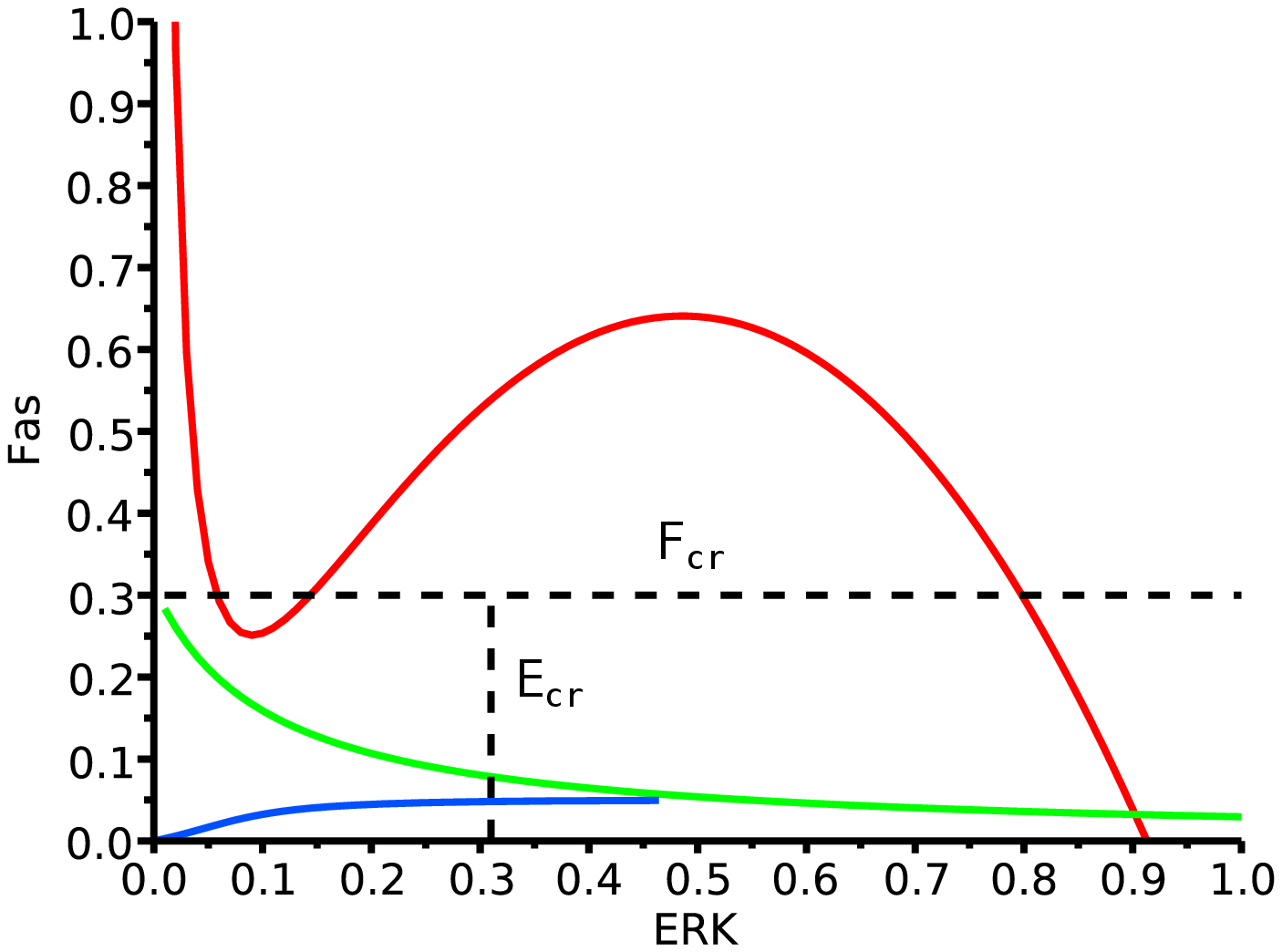}
\includegraphics[scale=0.43]{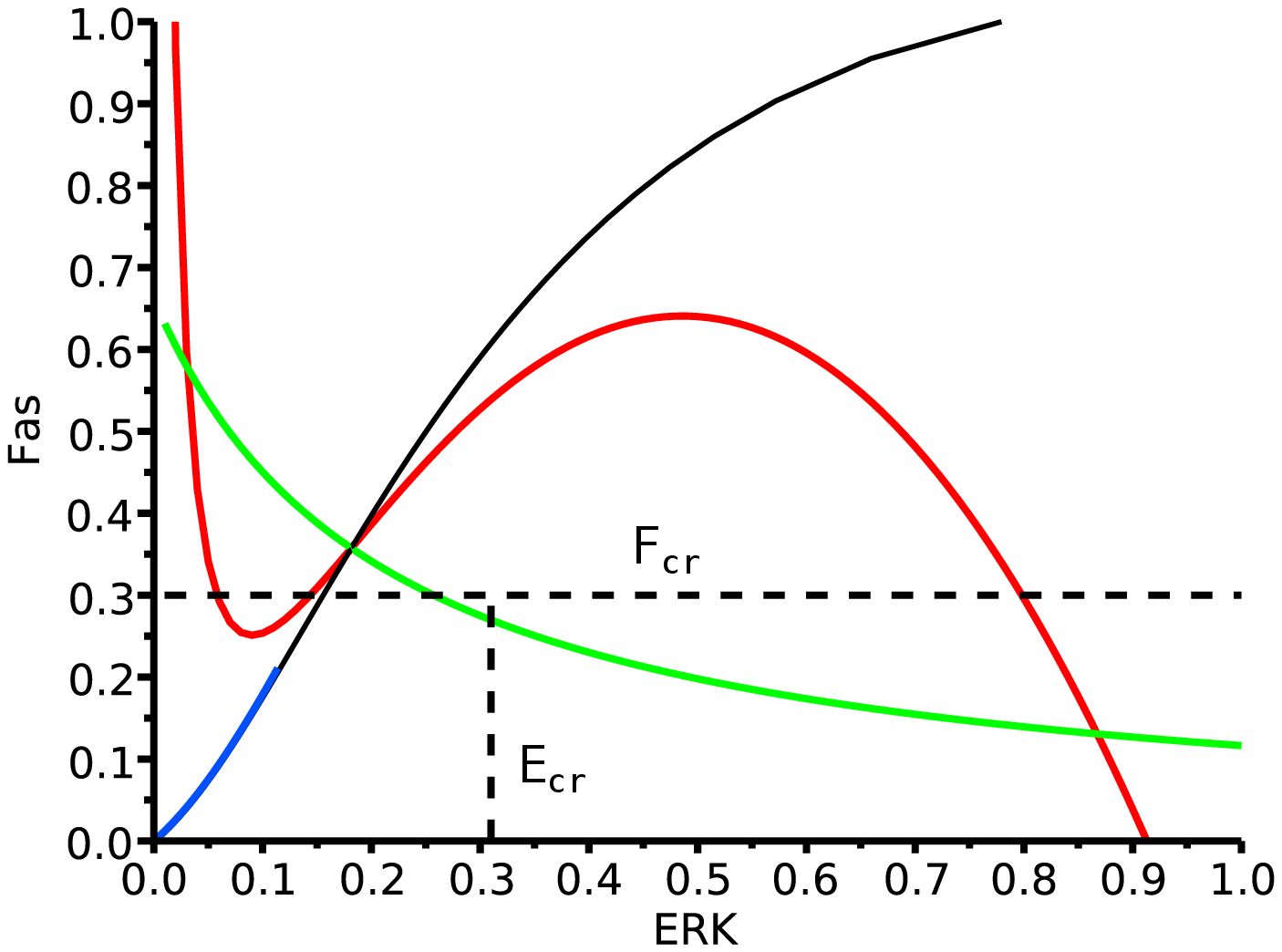}
\end{center}
A\hspace{6.7cm} B
\begin{center}
\includegraphics[scale=0.43]{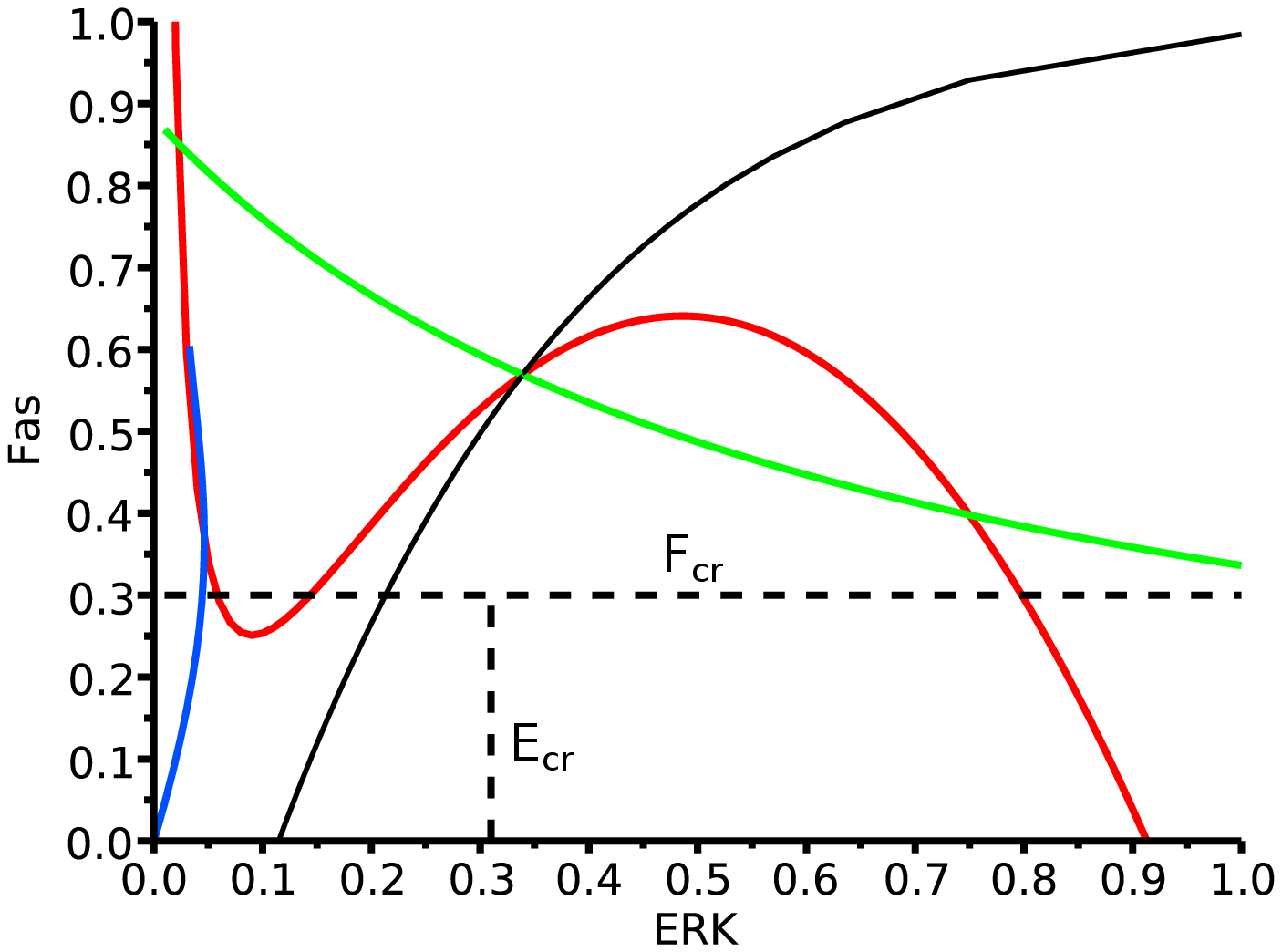}
\end{center}
\hspace{3.2cm} C
\caption{\textbf{Illustration of the `out of equilibrium' hypothesis.} All daughter cells are initially placed at the origin of the (Erk,Fas) plane and, depending on the external Fas-ligand value, try to escape the default differentiation zone (bottom left area) into apoptosis (above the dashed $F_{cr}$ line) or self-renewal zone (bottom right area) before the end of their cycle. Trajectories from the origin are illustrated by the straight blue curve. A: Low exposure to Fas-ligand leads to self-renewal. B: Medium exposure to Fas-ligand leads to differentiation. C: High exposure to Fas-ligand leads to apoptosis.}
  \label{sol1}
\end{figure}

In the case where the system is strictly monostable, this solution is nevertheless not satisfactory (no subpopulation appears because cells remain very close to each other). However, when the system is locally bistable, cells will converge either toward the lower stable branch or the upper stable branch of the $f$ zero-line. For a given value of Fas-ligand, the separatrix will cut the origin, so there is a clear threshold between the two states. Even if there is still a continuum of cells on the phase plane for continuous values of exposition to Fas-ligand, two denser groups appear. In fact, cells exposed to high Fas-ligand values quickly converge toward the top, cells exposed to low Fas-ligand values toward the bottom-right and those exposed to medium values remain trapped near the origin in a `slow' zone around the separatrix. The latter cells keep low Erk and Fas values, so by introducing the thresholds, three subpopulations are easily obtained (Figure \ref{sol1}). 

\begin{figure}[!ht]
 \centerline{\includegraphics[scale=0.75]{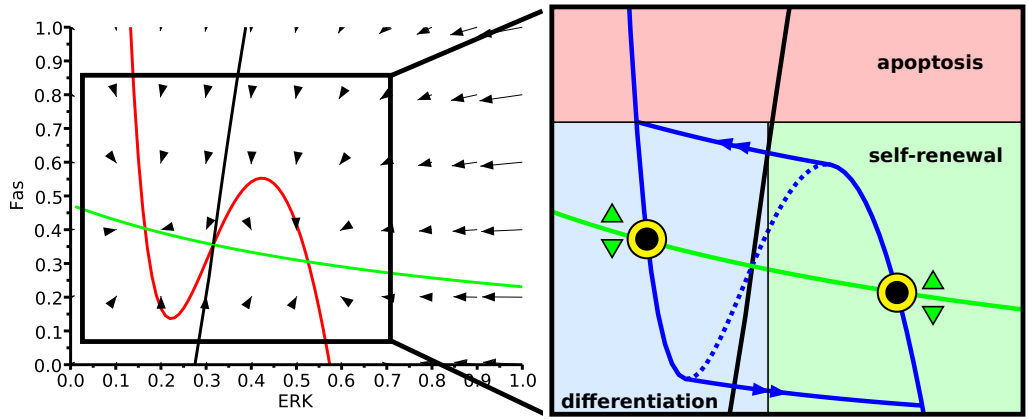}
}
 \caption{\textbf{The `hysteresis cycle' scenario.} Daughter cells initially inherit half of their mother's protein level, `choose' one of the stable branches depending on their exposition to Fas-ligand, and then move on these branches which form an hysteresis cycle, making the decision hardly reversible. Shown with yellow circles are the two possible positions for newborn cells exposed to exactly the same value of Fas-ligand on the hysteresis cycle. One can see how the initial choice is important: cells whose mother had low Fas values will be located on the lower branch, those whose mother had sufficiently high Fas values will switch to the upper branch.}
 \label{sol2}
\end{figure}

The second idea is that working at steady-state is more robust. When the system is locally bistable, one notices that cells on the lower stable branch are those which will self-renew (high values of Erk, reasonable values of Fas). If quantities of Erk and Fas are equally divided between the two daughter cells at birth, those exposed to medium values of Fas-ligand will appear near the separatrix and be divided between the upper branch (differentiation/apoptosis) and the lower branch (self-renewal). We thus have a system where cells rapidly take the decision between self-renewal or differentiation/apoptosis. Once the decision is made, cells are trapped on a hysteresis cycle controlled by Fas-ligand, so the decision is hardly reversible. Fas values have to become very high in order to get from the lower branch to the upper branch and conversely, very low in the opposite direction (Figure \ref{sol2}).

For this solution, called `hysteresis cycle case', introducing thresholds in order to determine cell fate is quite simple. The Erk threshold can be situated anywhere between the two branches and the Fas threshold somewhere on the upper branch, cutting into two the cells on this branch and so easily controlling the differentiation/apoptosis ratio.

The `out of equilibrium' and the `hysteresis cycle' scenarii are schematically described in Figure \ref{fig-two-scenarii}.

\subsubsection{Feedback Control Role}\label{feedback}

A key point in system (\ref{erk})--(\ref{fas}) is how it reacts to feedback controls. The analysis performed in the previous subsections dealt with the influence of each parameter on the system dynamics, mainly parameters related to feedback controls ($\alpha$ and $\gamma$). Let now study how parameters are affected by feedback controls. Erythropoietin (through function $\alpha$) is expected to decrease apoptosis and enhance self-renewal/differentiation. We also know that, at a smaller space scale, Fas-ligand induces apoptosis (through function $\gamma$) and other growth factors, on the opposite, stimulate self-renewal (through the function $\alpha$).

In order to validate possible dynamics, the effect of each factor must be checked in both cases, the `out of equilibrium' and the `hysteresis cycle' scenarii. 

The control exerted by Fas-ligand, which increases the value of Fas reached at steady-state, has already been largely discussed. Simply by existence of the threshold $F_{cr}$, Fas-ligand plays the expected role. Fas-ligand is linked to the parameter $\gamma$, which is actually a function of~$F_L$.

Epo is known to have a direct action on apoptosis \cite{kb1990}, independent of Erk pathway \cite{chappell-etal1997, nagata-etal1998, sui-etal2000}. Therefore, a simple idea is to increase $F_{cr}$ with the level of Epo. This clearly reduces apoptosis in favor of differentiation and self-renewal. However, it is also known that Epo, and other growth factors, activate Erk. So erythropoietin and growth factors are supposed to influence the variations of $\alpha$. It is noticeable that at a local scale, many growth factors are supposed to have similar mechanisms, so $\alpha$ should in fact be a function of many global and local growth factors.

The value of $\alpha$ controls the monostable/bistable bifurcation (Figure \ref{params}). When the system is locally bistable, $\alpha$ mainly controls the value of the local minimum of the $f$ zero-line. In the case where we work out of equilibrium, $\alpha$ changes the position of the separatrix and of the upper steady states, so when $\alpha$ is increased the separatrix cuts the origin for a higher value of $\gamma(F_L)$ and cells tend to be displaced toward steady-states with lower values of Fas and higher values of Erk. In the case where the hysteresis cycle is used, increasing the local minimum of $f$ reduces the size of the cycle. Cells located on the upper stable branch instantly `fall' on the lower stable branch, so self-renewal is immediately enhanced and differentiation reduced.

The effect of $\alpha$ is not the same in both dynamics, but finally it is clearly possible to decrease apoptosis in favor of self-renewal/differentiation and the trends for each protein is roughly what is expected biologically. However, it is still necessary to study whether the feedback is as efficient as observed experimentally, when the intracellular scale is coupled to the extracellular one (see Section~\ref{s-results}).


\subsection{Extracellular Scale} \label{computational}

In the previous section, we focused on the intracellular scale of the hybrid model, based on a system of ordinary differential equations describing competition between two key proteins, Erk and Fas. Concentrations of Erk and Fas induce either cell self-renewal, or differentiation, or death by apoptosis. We now focus on an other layer of the hybrid model, the discrete part, which describes how cells interact and how they influence intracellular parameters which depend on extracellular molecules.

In order to describe the evolution of an erythroid cell population, we chose to use an individual-based model. All cells and their interactions are then numerically computed. The objective being to represent erythroblastic islands, which have a limited size, such an approach takes automatically into account small population effects as well as random effects (direction of division, cell cycle length).

A population of cells is numerically simulated in a 2D computational domain which is a rectangle. Each cell is a discrete object, an elastic ball, considered to be circular and composed of two parts: a compressible part at the border and a hardly compressible part at the center. All newborn cells have the same radius $r_0$ and linearly increase in size until the end of their cycle, when they reach twice the initial radius. When a cell divides, it gives birth to two small cells side by side, the direction of division being randomly chosen. The cell cycle duration is itself variable. From a biological point of view, cell cycle proceeds through $G_0/G_1$, $S$, $G_2$ and $M$ phases. Duration of the first $G_0/G_1$ phase and transitions between phases are controlled by various intracellular and extracellular mechanisms, inducing stochasticity in cell cycle durations \cite{golubev, rew}. We assumed the duration of $G_0/G_1$ phase is a random variable with a uniform distribution in some given interval, other phase durations were supposed to be constant.

Under the assumption of small deformations, the force acting between cells can be expressed as a function of the distance between their centers. The force between two particles with centers at $x_i$ and $x_j$ is given by a function $D(d_{ij})$ of the distance $d_{ij}$ between the centers. This function is zero if the distance is greater than the sum of their radii. To describe the motion of a particle, one should determine the forces acting on it from all other particles and possibly from the surrounding medium. The motion of each cell is then described by the displacement of its center by Newton's second law:
\begin{equation}\label{odr}
m \ddot{x}+\mu m \dot{x} - \sum_{j\neq i} D (d_{ij})=0,
\end{equation}
where $m$ is the mass of the particle, the second term in left-hand side describes the friction by the surrounding medium, the third term is the potential force between cells. The force between two spherical particles is considered in the form
\begin{equation} \label{eql21}
D(d_{ij})=\left\{
 \begin{array}{ll}
K \displaystyle\frac{d_{0}-d_{ij}}{d_{ij}-d_{0}+2H_1},& d_{ij} < d_{0},\vspace{1ex}\\
0,& d_{ij}\geq d_{0},
 \end{array} \right.
 \end{equation}
where $d_0$ is the sum of cell radii, $K$ is a positive parameter, and $H_1$ accounts for the compressible part of each cell. The force between the particles tends to infinity when $d_{ij}$ decreases to $d_0-2H_1$. On the other hand, this force equals zero if $d_{ij} \geq d_0$.

Three cell types are computed. First, erythroblasts, which are immature erythroid cells, also known as erythroid progenitors. They follow the growth rules explained above and their fate is determined as described in Section \ref{mathematical}. They either self-renew and give two cells of the same type, or differentiate and give two differentiated cells (reticulocytes), or die by apoptosis, depending on their exposition to growth factors and Fas-ligand. It is noticeable that no asymmetric division is considered in this model. 

Second, reticulocytes. From a biological point of view, reticulocytes are almost mature red blood cells that leave the bone marrow and enter the bloodstream after ejecting their nuclei. In this individual-based model, they are differentiated cells which stay in the bone marrow a little while after being produced, and leave the bone marrow (computational domain) after a time equal to one cell cycle duration. Contrary to erythroblasts, reticulocytes do not have a choice to make, they only express Fas-ligand, thus influencing the development of surrounding erythroblasts.

Third, macrophages. They are big white blood cells located at the center of erythroblastic islands. They very probably have an active role in the development of blood cells surrounding them. In this model, they express growth factors, driving nearby erythroblasts toward differentiation and self-renewal.

Macrophages express growth factors that are supposed to diffuse in their neighborhood. Reticulocytes express Fas-ligand on their surfaces which does not diffuse in a real bone marrow, yet the expression of Fas-ligand is modeled by short-diffusion. This allows considering several phenomena that are not taken into account in the hybrid model: first, cell motion in the bone marrow is actually more important than computed and, second, there are normally several maturing erythroid subpopulations which increasingly express Fas-ligand, creating a gradient of exposition. Moreover, cells are not circular in reality and this short-range diffusion compensates their inexact representation. Both Fas-ligand ($F_L$) and growth factor ($GF$) concentration evolutions are described with the following reaction-diffusion equations,
\begin{eqnarray}
\frac{\partial F_L}{\partial t} &=& D_{F_L} \Delta F_L + W_{F_L} - \sigma_{F_L} F_L, \label{fas-ligand} \vspace{1ex}\\
\frac{\partial GF}{\partial t} &=& D_{GF} \Delta GF + W_{GF} - \sigma_{GF} GF,\label{gf}
\end{eqnarray}
where $W_{F_L} = k_{F_L} C_{ret}$ is a source term depending on the number of reticulocytes ($C_{ret}$ denotes the relative volume of reticulocytes in the computational domain) and $W_{GF}$ is a constant source term for growth factor (released by the macrophage) concentration, $\sigma_{F_L}$ and $\sigma_{GF}$ are degradation rates, and $D_{F_L}$ and $D_{GF}$ are diffusion rates. If the diffusion coefficient $D_{F_L}$ is sufficiently small, then Fas-ligand is concentrated in a small vicinity of reticulocytes. In this case, Fas-ligand influences erythroid progenitors when they are sufficiently close to reticulocytes. This is the short-diffusion illustrated in this work.

\begin{figure}[!ht]
 \centerline{
 \includegraphics[scale=0.35]{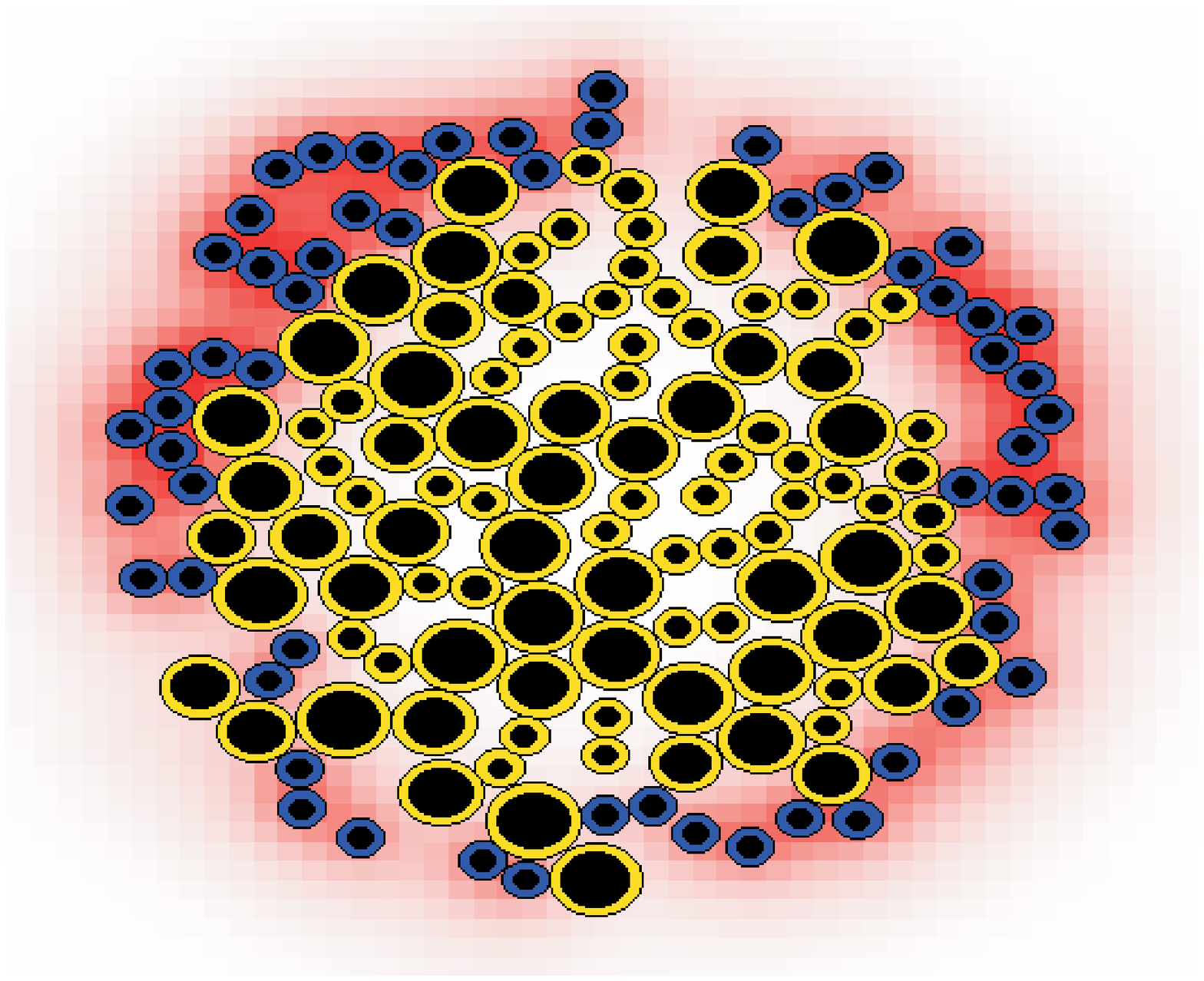} \includegraphics[scale=0.43]{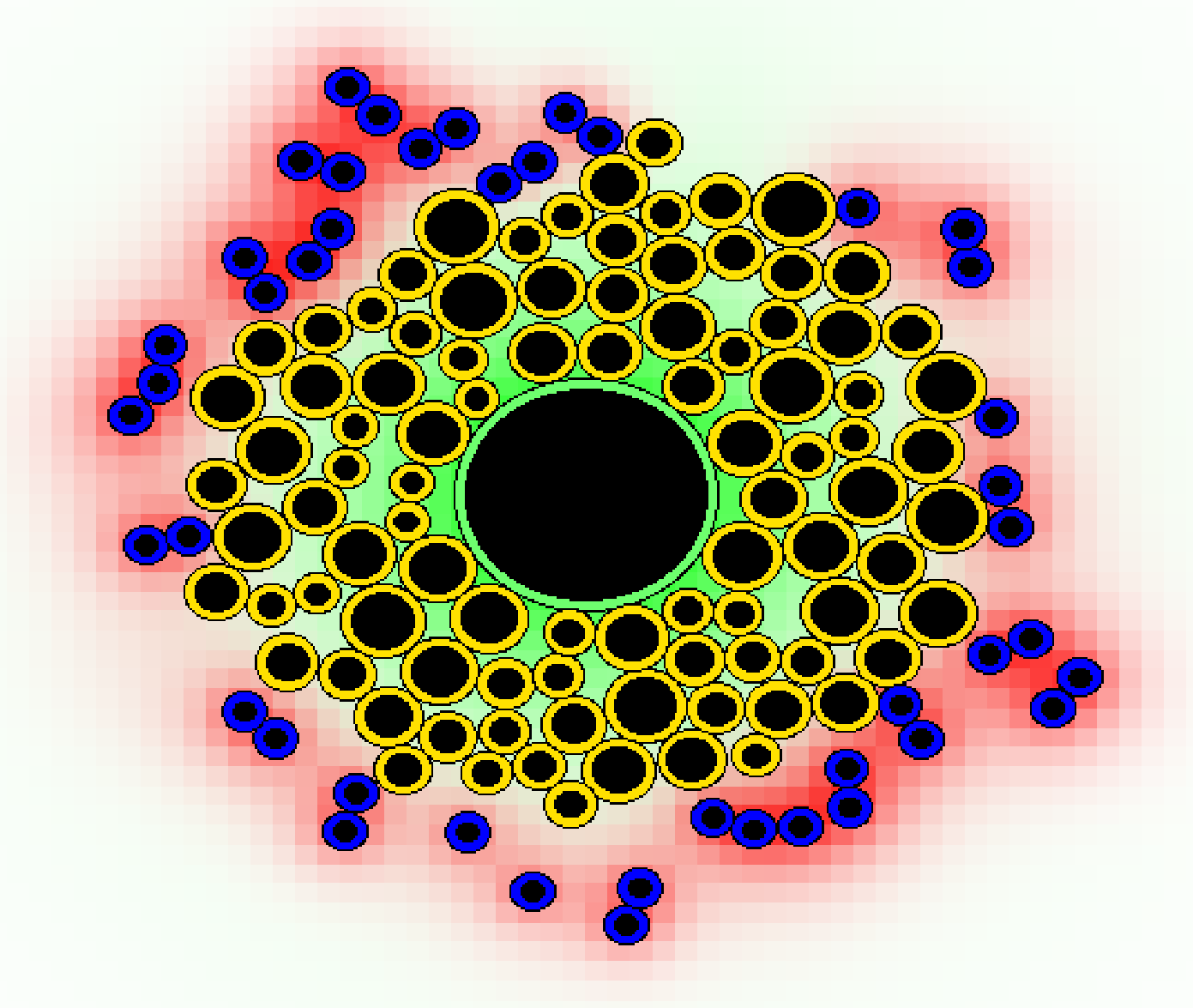}
 }
 \caption{\textbf{Screenshots of the software program with and without macrophage. }The green cell is a macrophage, yellow cells are immature cells (erythroblasts) and blue cells are differentiated cells (reticulocytes). The green substance is secreted by the macrophage (growth factors) and the red substance by reticulocytes (Fas-ligand). In each cell, the black area represents the incompressible part of the cell.}
 \label{program}
\end{figure}

Figure \ref{program} shows an example of erythroblastic island in the hybrid model. The big green cell in the center is a macrophage, the green substance surrounding it is the above mentioned growth factors released by the macrophage. At the border, blue cells are reticulocytes emitting Fas-ligand (whose concentration is represented in red). Between them are the erythroblasts, growing in size until they die or divide into two new cells, depending on how they were influenced by the two diffusing proteins.

A fourth cell type, erythrocytes, is considered in this model. Biologically, erythrocytes are mature reticulocytes, circulating in blood and transporting oxygen to the organs. In the model, erythrocytes are reticulocytes that have spent a time equal to one cell cycle in the computational domain and then left it. Hence, erythrocytes are only counted as cells outside the domain (they are not represented on the computational domain) which act, through feedback loops, on the fate of immature cells within the domain. We will assimilate erythrocytes and red blood cells (RBC) in the following, by considering that RBCs are only erythrocytes (in reality, some reticulocytes usually manage to leave the bone marrow and enter the bloodstream where they end their maturation, so they are also red blood cells). Erythrocytes will be supposed to survive (arbitrarily) during a time corresponding to 3 cell cycle durations (although RBCs do not cycle). This is short, since in mice RBCs have a lifespan of 40 days, yet it allows performing a first feedback test and it can be changed when comparing the model to experimental data.

\subsection{Coupling Both Scales}

Now that both modeling scales have been described and a general picture of the model is accessible, it becomes necessary to detail how the two scales can be linked together. At the highest level, there is a small erythroblastic island formed with two populations of developping red blood cells and, optionally, a central macrophage. The island has a particular topology: the macrophage is at the center, differentiated cells (reticulocytes) at the border and immature cells in between. Both macrophage and differentiated cells emit growth factors continuously controlling intracellular protein concentrations: immature cells at the center are rather exposed to growth factors produced by the macrophage whereas those at the border are rather exposed to Fas-ligand. In addition, immature cells are subject to a feedback control mediated by mature red blood cells circulating in the bloodstream, representing the action of erythropoietin. Concentration of Epo in the computational domain is supposed to be uniform, so all cells are similarly influenced by Epo. This assumption holds true for all external substances acting on cells at a global scale, yet other more sophisticated choices are possible.

Erythroid progenitors are then exposed to a continuum of erythropoietin $Epo$, growth factors $GF$ and Fas-ligand $F_L$. Regarding the intracellular scale, this means that $\alpha$ and $\gamma$ take continuous values (see (\ref{erk})--(\ref{fas})). Let us remind that high $\alpha$ drives erythroblasts to self-renewal, high $\gamma$ to death, and intermediate values to differentiation. Hence, the cell population scale influences the intracellular scale through Epo, growth factors and Fas-ligand concentrations, that is through functions $\alpha$ and $\gamma$ and the critical value $F_{cr}$ of Fas concentration (Epo increases $F_{cr}$ in order to inhibit cell apoptosis).

One can notice that when there is no macrophage, there is only one value of $\alpha$ for all erythroblasts (this value depends on the concentration of Epo which is related to the number of erythrocytes, and is supposed to be the same for all immature cells), so cells are simply situated along a unique $f$ zero-line. It is then convenient to investigate how the value of $\gamma$ will determine a cell's choice (see Section \ref{feedback}).

As just mentioned, Epo-mediated feedback control simply corrects the critical value for apoptosis $F_{cr}$ and values of $\alpha$ uniformly, so its impact can be easily understood individually by referring to the analysis in Section \ref{feedback}. However, in order to compute this feedback control, the number of erythrocytes in blood circulation must be estimated. This is performed from the reticulocyte population as mentioned at the end of the previous section.


\section{Results: Stability of Erythroblastic Island and Function of Central Macrophage}\label{s-results}

In the previous section, the hybrid model has been presented in details. The applicability of such a modeling will be investigated in this part. To do so, we focus on finding conditions (parameter values, that are partially discussed in Appendix \ref{parameters}) for the system leading to stable erythroblastic islands (this means the number of cells in the island stays almost constant for a certain time when not facing a perturbation), reacting sensibly to global feedback variables (for example in the case of anemia) and proving to be robust to parameter values variations. 

We begin by investigating the case of one erythroblastic island without macrophage in its center. The case of the island with a macrophage is then considered to see how the stability of the island is affected. For these analyses, simple numerical stability and feedback tests were performed to check the validity of hypotheses.

\subsection{Erythroblastic Island without Macrophage}

Consider an island initially composed of 98 erythroid progenitors and 64 reticulocytes surrounding them. Parameters are carefully chosen to optimize stability and response to feedback (see Appendix \ref{parameters}). 

\subsubsection{Stability Analysis}\label{ss-stability}

Erythroblastic islands are likely to be stable. Indeed, one can expect that, in order to be biologically responsive, an island survives for several cell cycles. It is relevant to determine whether it is mathematically possible to achieve long-lasting stability with the hybrid model or not.

\begin{figure}[!ht]
\centerline{
\includegraphics[scale=0.36]{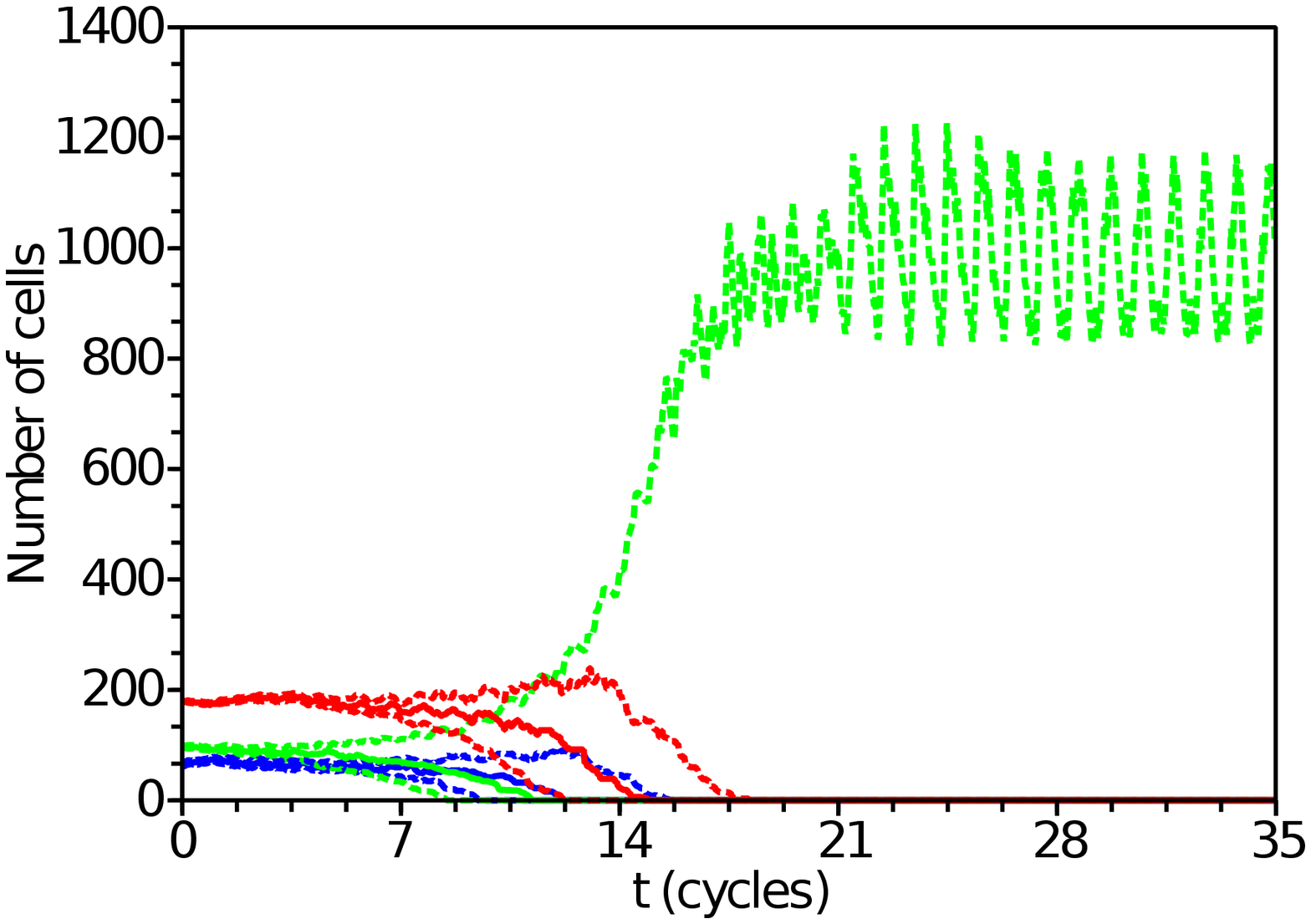} \includegraphics[scale=0.36]{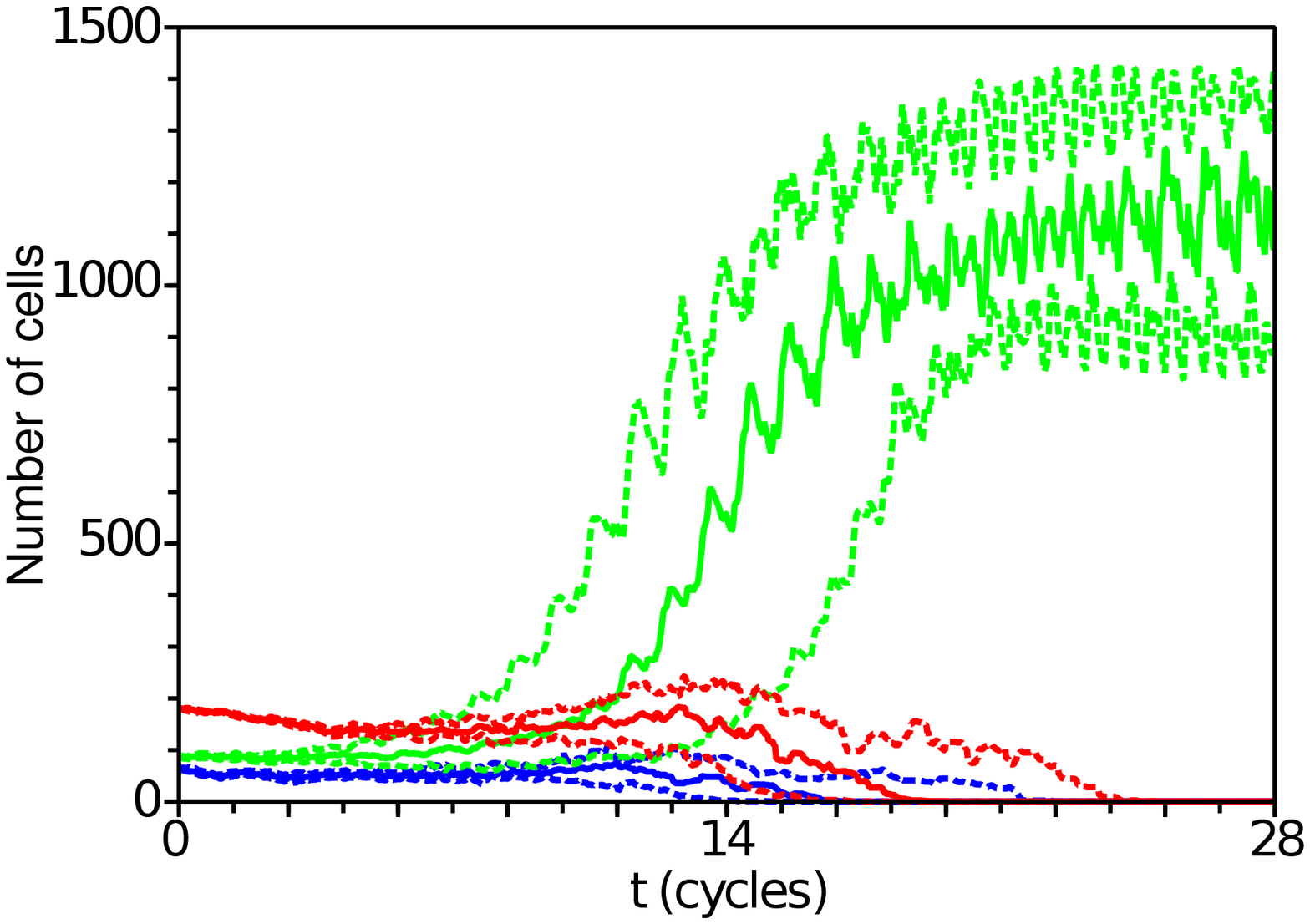}
}
 \caption{\textbf{Illustration of the fate of the most stable erythroblastic islands in the absence of macrophage.} Results were obtained out of equilibrium (left) and using the hysteresis cycle (right). Green lines represent the number of erythroblasts, blue lines reticulocytes, red lines estimation of circulating red blood cells produced by the island (thick lines are medians, thin lines quartiles over 40 simulations). Observed saturation is only due to space limitation.} 
 \label{sistability}
\end{figure}

A wide range of parameters has been tested for the two possible dynamics, that is either out of equilibrium or with the hysteresis cycle. At first, the best results look like quite similar (Figure \ref{sistability}). In both cases, the island remains approximately stable during some cycles but, due to stochastic variations (cell cycle duration, orientation of the mitotic spindle at division, small size populations), it suddenly dies out or excessively proliferates (it must be noticed that saturation is only due to space limitations). Two questions arise from this fact: why does the system behave this way and is it still biologically relevant?

Let's focus first on the mechanisms leading to an island's stability. In the simulations, during each cycle there is a turnover of reticulocytes, in order to replace those which entered bloodstream. These reticulocytes then induce apoptosis and differentiation in the erythroblast population. Within this population, in order to remain stable, the proportion of self-renewing erythroblasts has to be constant. From a geometrical point of view, self-renewing erythroblasts are located at the center of the erythroblastic island, differentiating and dying erythroblasts at the border (Figure \ref{program}). When a random variation occurs in the size of the island, the amount of the differentiating and dying cells varies like the perimeter of the circle (occupied by reticulocytes which act at constant range), while the amount of self-renewing cells varies like the area of the circle. As a result, the ratio `differentiating and dying cells' over `self-renewing cells' does not remain constant. When it derives too far from the initial value, one subpopulation rapidly overgrows the other.

It appears then that it is impossible to find parameter values for which an island would remain at steady-state on a long term. However, this does not imply that results are  biologically irrelevant. The fact that islands die in this model is not really surprising since, biologically, there might be a turnover of these structures and erythroid progenitors permanently arise from the differentiation of hematopoietic stem cells, which has been neglected here. Also, proliferation is not as dramatic as it might seem. First, it is possible to turn it down by varying parameters (but sacrificing instead the average lifetime of an island) and also proliferation may not be that simple in a realistic environment. Indeed, there is no obstacle in the computational domain and some particular geometries could clearly reduce variations in the proportion of self-renewing cells. Another point is that when several islands are side-by-side, proliferating islands will collide and be in contact with more reticulocytes than before, which may block their growth (data not shown). Nevertheless, unstable islands revealed hard to control and stability of islands should be expected, as previously mentioned.

Regarding the two possible dynamics, either out of equilibrium or hysteresis cycle, stability of the system at equilibrium (hysteresis cycle) is slightly better (Figure \ref{sistability} right). Moreover, a significant fact arose during experiments when distribution of cell cycle lengths was varied: the dynamics at equilibrium is very robust in this case, whereas the other dynamics becomes clearly biased. Indeed, when cells are placed at the origin at the beginning of their cycle, they start in the differentiating area. When their cell cycle length is short, they do not have enough time to escape this area and differentiate. When cell cycle duration is long, they cannot differentiate. Consequently, some cell fates are only decided because of cell cycle length, independently of the position in the erythroblastic island, and the island structure is quickly lost. In conclusion, since cell cycle length variation may be large \cite{golubev,rew}, it was decided to abandon this `out of equilibrium' dynamics and to perform further tests with the dynamics using the hysteresis cycle (Figure \ref{fig-two-scenarii}, right panel).

\subsubsection{Feedback Relevance and Relation to Stability}\label{sss-feedback}

The above-described hybrid system should be clearly influenced by feedback control loops, as expected biologically, and, besides, feedback loops may play a role in the stabilization of erythroblastic islands.

As mentioned in Section \ref{mathematical}, different feedback controls are considered in this model: a local control through Fas-ligand, activating Fas and influencing erythroblast differentiation and death by apoptosis; two global feedback controls by erythropoietin, one activating Erk through the function $\alpha$, the second one varying the critical value of Fas, namely $F_{cr}$, inducing apoptosis. We focus here on the global feedback control mediated by Epo.

\begin{figure}[!ht]
\centerline{
\includegraphics[scale=0.37]{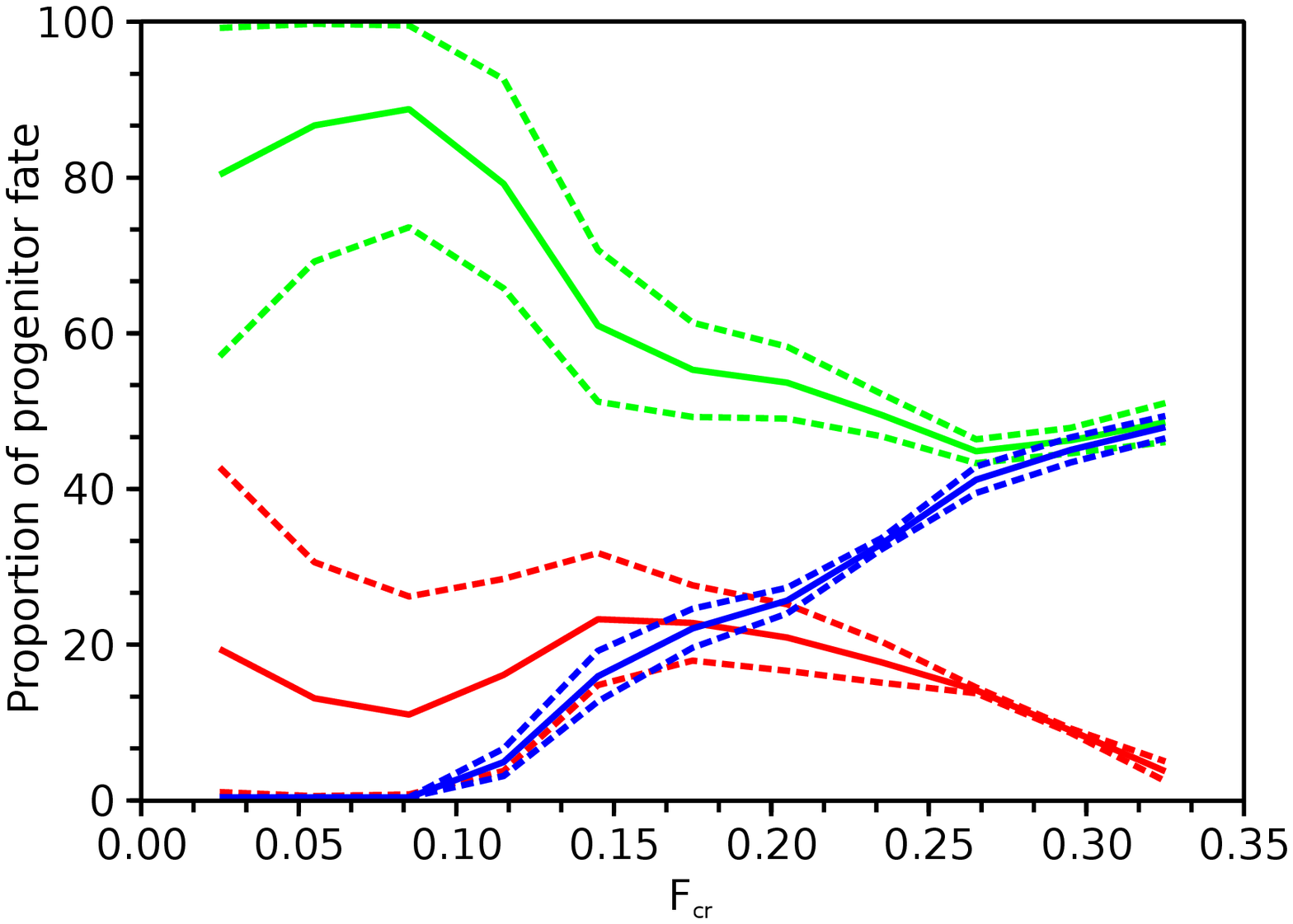} \includegraphics[scale=0.37]{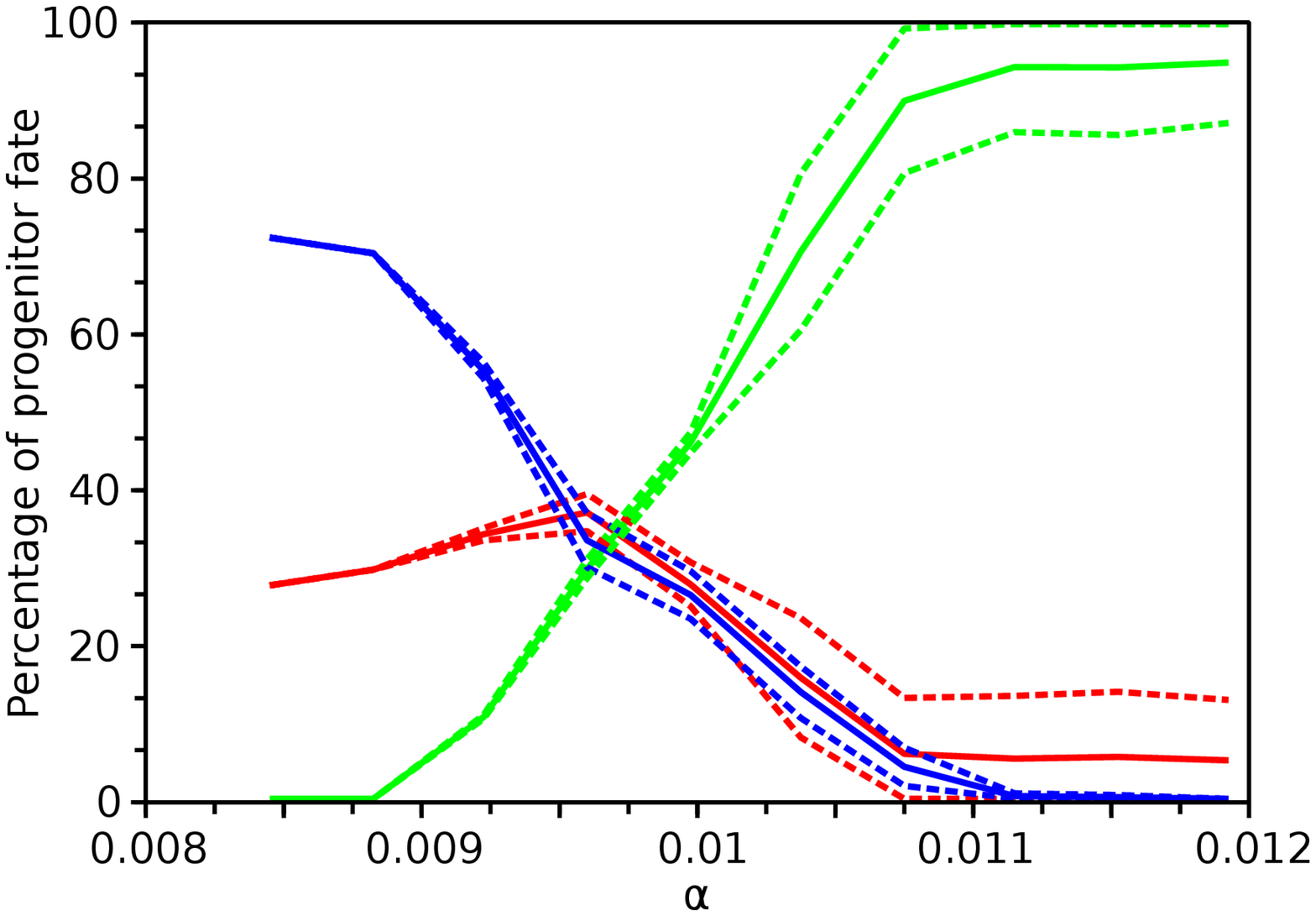}
}
\caption{\textbf{Effect of feedback controls on progenitor subpopulations in the absence of macrophage in the island.} Each curve represents the percentage (means $\pm$ standard deviation) of self-renewing (green), differentiating (blue) or apoptotic (red) erythroblasts during the first five cell cycles, as function of $F_{cr}$ (left panel) and $\alpha$ (right panel). }
 \label{sifeedback}
\end{figure}

It is known that increasing Epo levels in the bloodstream, and consequently in the bone marrow, increases Erk activity and decreases erythroblast apoptosis rate. Hence, an increase of Epo levels induces an increase in the values of both $\alpha$ and $F_{cr}$. The exact influence of Epo on these two quantities is however not known. Therefore, we began by simply monitoring the behaviour of one island under constant values of these two parameters around the default value (Figure \ref{sifeedback}).

By comparing the proportions of different erythroblast subpopulations (self-renewing, differentiating and apoptotic populations) during the first 5 cycles, it appeared that an increase of $F_{cr}$ (that is, an increase of Epo levels) decreases apoptosis and self-renewal in favor of differentiation. For low values of $F_{cr}$, there is no differentiation (erythroblasts at the center of the island survive, others die), so the island proliferates by absence of death factors. For higher values, effect on self-renewal can be tuned by modifying parameters controlling the hysteresis cycle. At some point the self-renewing population is not affected anymore and there is simply a transfer from the apoptotic subpopulation to the differentiating subpopulation (the differentiation part of the differentiation/apoptosis branch increases).

Increasing $\alpha$, on the contrary, mainly acts on the self-renewing population. An increase of $\alpha$ sends all cells on the self-renewing branch of the hysteresis cycle and differentiation/apoptosis disappears. Reducing $\alpha$ increases the size of the apoptosis/differentiation branch. For a constant value of $F_{cr}$, only the differentiation part of this branch is increased. One thus observes a transfer from the self-renewing branch to the differentiation part of the other branch.

In conclusion, the system's response is qualitatively what is expected. Epo mainly decreases apoptosis in favor of differentiation and stimulates self-renewal. However, the statistics computed above only indicate trends and it is important to see how islands actually react to this feedback.

The next step is then to explicitly include feedback loops in the system's equations to see whether the response is quantitatively correct or not, and if it is possible, through feedback controls, to achieve a better stability than before. Therefore, we assumed a linear variation of both $\alpha$ and $F_{cr}$ with deficiency of red blood cells (RBC), hence implicitly describing the dependency on Epo. Denoting by $N_{RBC}$ the number of circulating red blood cells, and since a decrease of RBC count induces an increase in Epo levels, one gets
\begin{equation}
\label{alphafeedback}
\alpha := \alpha(N_{RBC}) = {\alpha}_0 + k_{\alpha} (N_{target} - N_{RBC}),
\end{equation}
\begin{equation}
\label{fascrfeedback}
F_{cr} := F_{cr} (N_{RBC}) = F_0 + k_{F} (N_{target} - N_{RBC}).
\end{equation}
We remind that $N_{RBC}$ is estimated by the number of reticulocytes which have left bone marrow and are supposed to survive during a time equivalent to 3 cell cycles. $N_{target}$ is the number of RBCs in circulation during a typical run before proliferation/extinction of the island. Parameters ${\alpha}_0$ and $F_0$ are `typical' values of $\alpha$ and $F_{cr}$ (as given in Appendix \ref{parameters}). Parameters $k_{\alpha}$ and $k_{F}$ are to be estimated.

It must be mentioned that these two feedback controls are not exact from a biological point of view, since one would expect the functions to saturate for low and high values of $N_{RBC}$, yet they are good approximations when the number of red blood cells is not too far away from the target (for instance, in normal erythropoiesis).

Functions $\alpha$ and $F_{cr}$ in (\ref{alphafeedback}) and (\ref{fascrfeedback}) were used together with the previous initial conditions (hysteresis cycle case), and it was impossible to achieve better stability than previously mentioned when varying the values of $k_{\alpha}$ and $k_{F}$ (data not shown). To be more precise, it was possible with strong feedback controls to stop proliferation, instead the island systematically died out. Other tests, based on simulations of a constant number of red blood cells ($N_{RBC}$), showed that the effect of feedback controls was not fast enough to deal with the quick proliferation and extinction of erythroblastic islands. The island still lived around 10--15 cycles and, when it began to expand or die out, feedback response came too late or was too strong to bring it back to its ideal size. Therefore, what was observed was only the usual destabilization of the island, feedback being only able to turn proliferation into extinction after some cycles.

As a result, it is very difficult to test feedback reliably on unstable islands. However, the study showed that global feedback loops are inefficient when it comes to controlling local structures, since proliferation or extinction events occur quickly in the model. 

In the next section, we consider the addition of a macrophage in the center of the island on the stability of the island and the relevance of feedback controls.

\subsection{Island with macrophage}

In this case, an erythroblastic island is initially composed of one macrophage, 80 erythroid progenitors and 64 reticulocytes (see Figure \ref{program}, right). As done in the previous section, stability of the island and influences of feedback controls are investigated. 

In addition to feedback controls exerted, in the previous section, on erythroblast dynamics by Epo and Fas-ligand, the macrophage in the center of the island is supposed to release growth factors, whose concentration is denoted by $GF$, which positively act on Erk activation. A number of proteins associated with the proliferation of erythroid progenitors and that could fulfill such a function have been described. This includes Stem Cell factor (SCF), the c-kit ligand, Ephrin-2, the EphB4 ligand, and BMP4, a member of the TGF$\beta$ family (\cite{rhodes2008}  and references therein). All those growth factors are assumed to diffuse around the macrophage as developed in Section \ref{computational} (Equation (\ref{gf})). They trigger erythroblast self-renewal through the function $\alpha$: the more growth factors, the higher $\alpha$. Function $\alpha$ now becomes
\begin{equation}
\label{alphagf}
\alpha := \alpha(N_{RBC},GF) = {\alpha}_0 + k_{\alpha} (N_{target} - N_{RBC}) + k_{GF} GF.
\end{equation}
Meaning of parameters $\alpha_0$, $k_{\alpha}$ and $N_{target}$ is identical to the ones defined for (\ref{alphafeedback}). Values are also the same, except for $\alpha_0$ that had to be decreased to compensate the addition of the term $k_{GF} GF$. Parameter $k_{GF}$ was set accordingly (influence of these parameters are discussed below). Default values are $\alpha_0=0 \: h^{-1}$ and $k_{GF} = 3 \: h^{-1}.\textrm{molecule}^{-1}$.

\subsubsection{Stability Analysis}

\begin{figure}[!ht]
 \centerline{\includegraphics[scale=0.6]{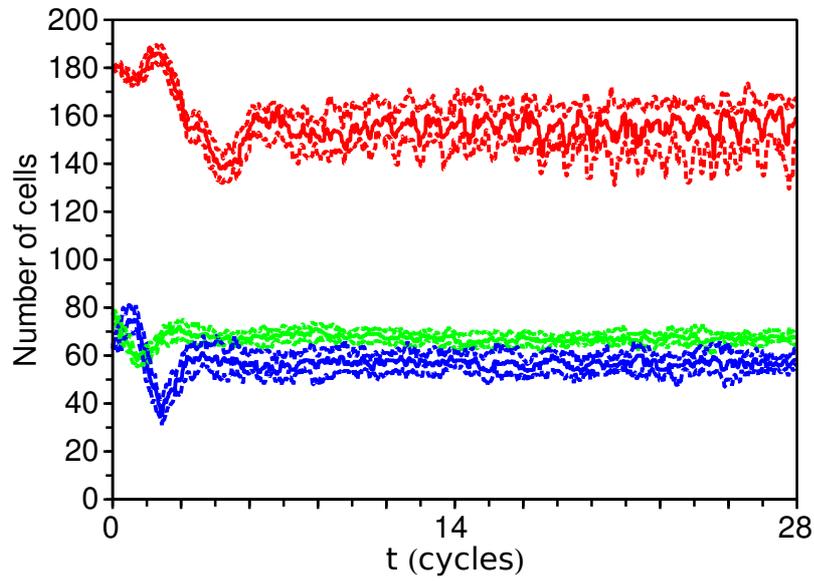}}
 \caption{\textbf{Number of erythroid cells in the presence of a macrophage in the center of the island.} Results were obtained using the hysteresis cycle. Green lines represent the number of erythroblasts, blue lines reticulocytes, red lines estimation of circulating red blood cells produced by the island (thick lines are medians, thin lines quartiles over 40 simulations). After addition of a macrophage, islands automatically stabilize for previously chosen parameters.}
 \label{macrostability}
\end{figure}

Stability of the hybrid system is first investigated without considering global feedback controls (that is, $k_{\alpha}=0$ in (\ref{alphagf}) and $k_{F}=0$ in (\ref{fascrfeedback})). Results are illustrated in Figure \ref{macrostability}.

For a large range of parameter values, the system quickly converges toward a steady-state. The number of self-renewing and differentiating cells (and thus circulating RBC) can be precisely controled by parameter values, as will be detailed in the next subsection. There is a first phase where the size of subpopulations oscillates (because of lack or abundance of death factors) yet steady-state is always quickly reached (within a few cycles), even though inherent stochastic oscillations are always present.

The macrophage completely controls the island. If ${\alpha}_0$ is sufficiently low, a cell will not be able to self-renew without external growth factors. Therefore, the default behaviour of an isolated cell is differentiation. On the other hand, a cell in contact with the macrophage will be exposed to a very high concentration of growth factors and will always self-renew, since it is generally not exposed to death factors. In the vicinity of the macrophage, there is a competition between growth factors and death factors, which will eventually determine the size of the island. If $\alpha_0$ or $k_{GF}$ gets higher, growth factors will reach further erythroblasts, thus the size of the final island will increase.

\subsubsection{Feedback Relevance and Relation to Stability}

Now that erythroblastic islands are stable, due to the presence of the macrophage, it is not necessary to be limited to a period of 5 cycles to compare the subpopulations of the island, as was done in the previous case (Section \ref{sss-feedback}). Rather, it becomes relevant to see how the size of the island and the production of RBC could evolve on a long term.

\begin{figure}[!ht]
 \centerline{\includegraphics[scale=0.37]{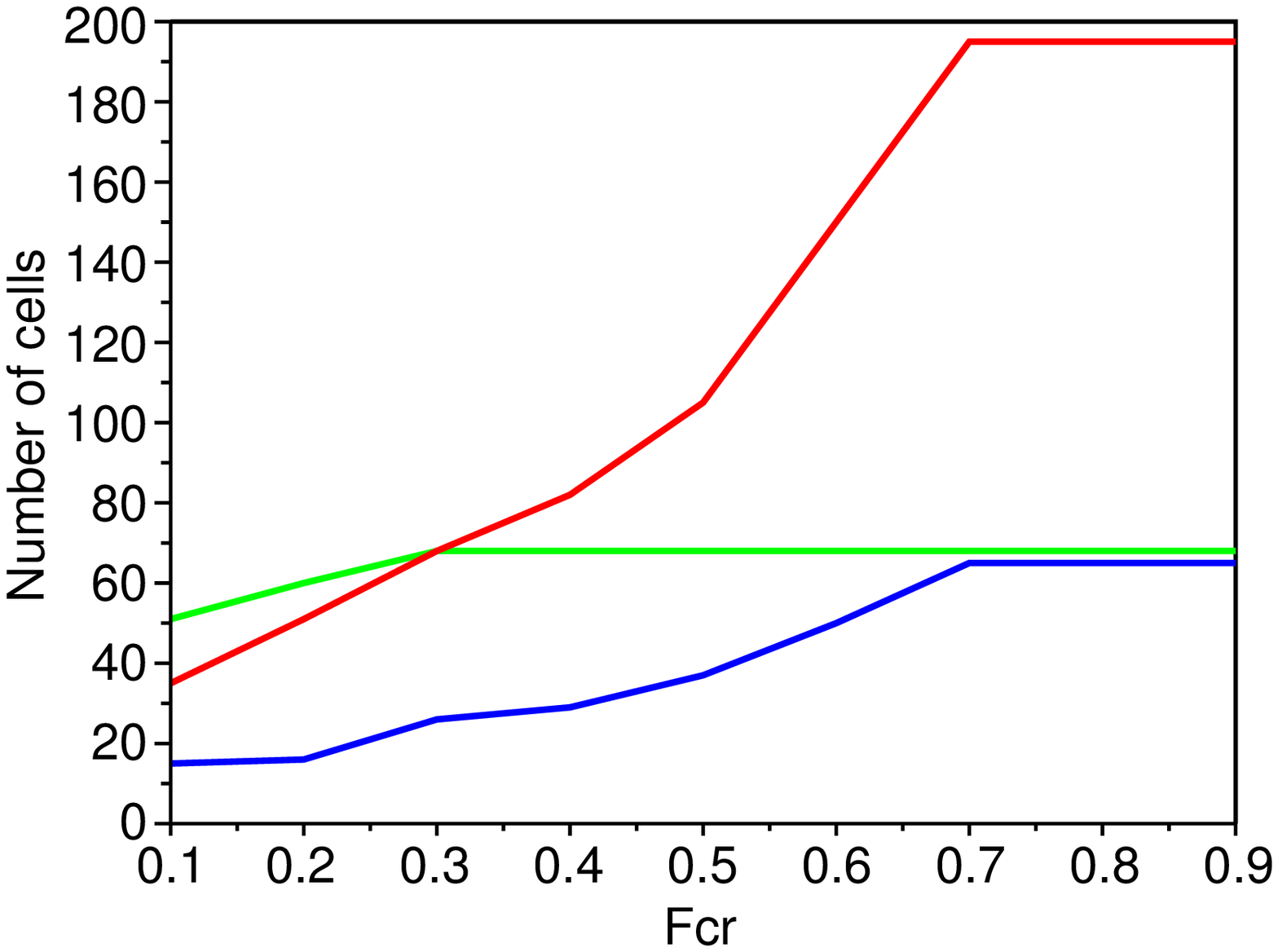}
 \includegraphics[scale=0.37]{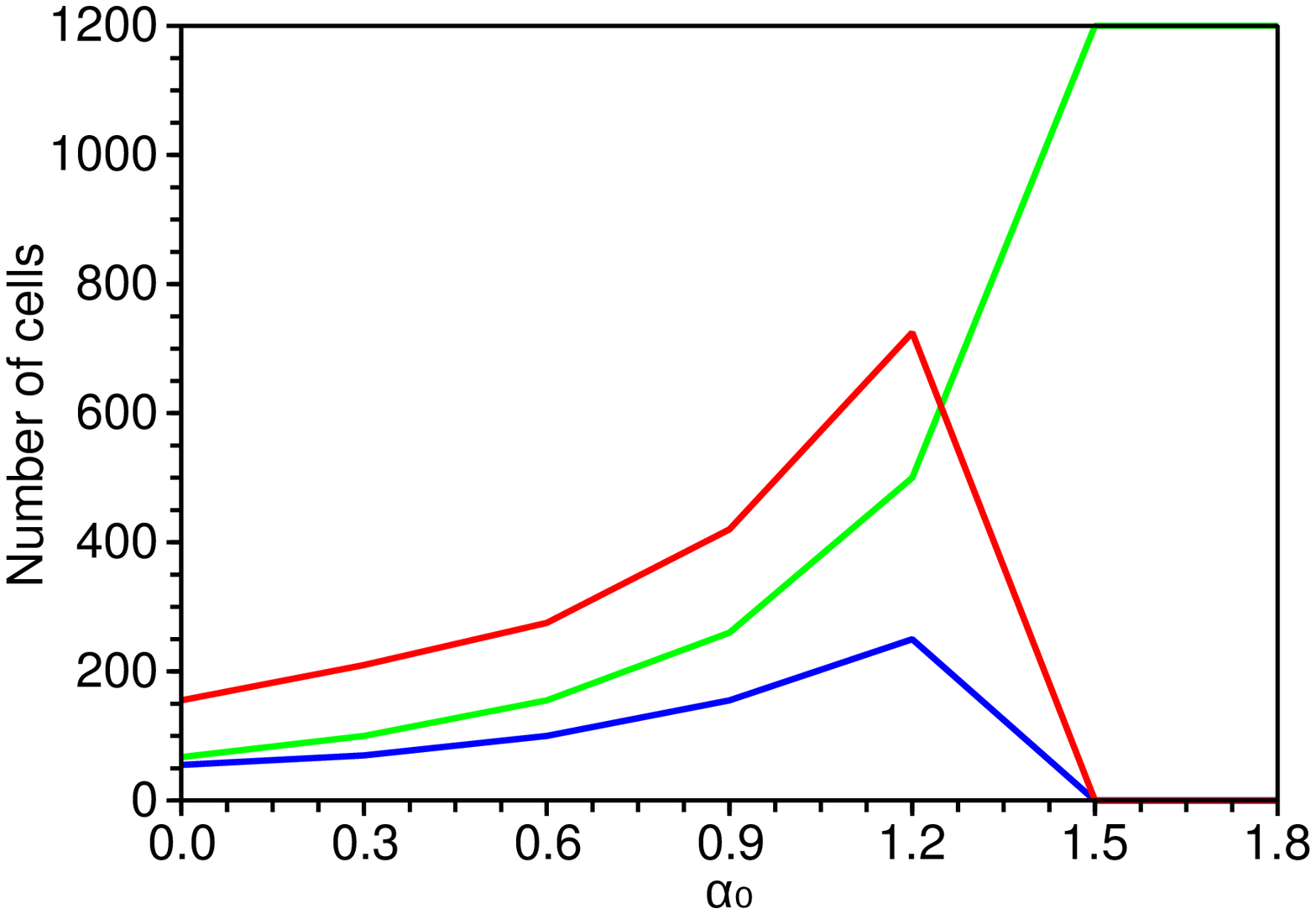}}
 \caption{\textbf{Impact of feedback controls on steady-state values when the erythroblastic island contains a macrophage.} Left panel: Influence of $F_{cr}$ variations; Right panel: Influence of $\alpha$ variations. Green lines represent erythroblast counts, blue lines reticulocyte counts and red lines RBC counts.}
 \label{macrofeedback}
\end{figure}

Similarly to what has been done in the previous case, we once again simulated the expected consequences (in terms of steady state value) of feedback control by running simulations under constant values of $\alpha$ and $F_{cr}$. Results are shown in Figure \ref{macrofeedback}.

When $F_{cr}$ increases, both the number of reticulocytes and RBC increases, as the differentiation part of the differentiation/apoptosis branch increases. For high values of $F_{cr}$, there is no apoptosis anymore. The number of erythroid progenitors seems to be rather independent of the value of $F_{cr}$.

When $\alpha$ increases, which can be seen as an increase of the ground level of global `surviving factors' (here Epo and macrophage-emitted growth factors), the size of the island increases mainly due to the increase of the self-renewing population. As a result, reticulocytes, situated all around the island, also increase in number, as does the amount of RBC. However, for very high values of $\alpha$, surviving signals are so strong that almost all erythroblasts self-renew (the macrophage has no effect anymore) and the island is made only of proliferating cells, reticulocyte and RBC counts vanish. Such high values of $\alpha$ may only be reached in extreme stress situations and mainly create a pool of cells which may afterwards differentiate and substantially increase the number of RBC at once (but this implies that another mechanism has to lower the value of $\alpha$ at some point). In any case, these values of $\alpha$ may else be avoided by adding a saturation effect on the feedback loop.

In each case, a substantial increase of RBC production (about 3-fold and 4-fold respectively) is observed. As the two mechanisms are largely independent, they can be combined and thus the overall production rate can be greatly increased, in magnitudes comparable to what can be observed biologically~\cite{chasis2008, rhodes2008, socolovsky2007}. This seems to indicate that the reaction of the system with macrophage to feedback controls is more efficient, relevant and easy to observe compared to the case without macrophage, again in good agreement with biological evidences~\cite{rhodes2008}.


\section{Discussion}

We proposed a new model for erythropoiesis modeling based on the coupling of two relevant scales, the intracellular scale consisting of protein regulatory networks and the extracellular scale focusing on cell movement and interactions as well as growth factor distribution in the medium. Description of the competition between two proteins within erythroid progenitors has been performed with continuous models (ordinary differential equations) whereas cells have been studied as discrete objects on an off-lattice model, so the whole system can be named `hybrid model', for it consists in the coupling of continuous and discrete systems. Applied to normal erythropoiesis in the bone marrow, in particular to the regulation of erythroblastic islands, the model suggests an important role of macrophages in the stability of islands. In the absence of macrophages, erythroblastic islands very quickly lose their stability, meaning they either die out or abnormally proliferate (with overproduction of erythroid progenitors). They survive only for the equivalent of about ten cell cycles. On the contrary, when the erythroblastic island is built around a central macrophage, assumed to release growth factors sustaining erythroid progenitor survival and proliferation (like SCF, Ephrin-2 or BMP-4 \cite{rhodes2008}), then the island is a very robust structure, able to produce continuously erythrocytes while keeping a steady structure (that is an almost constant ratio of reticulocytes over progenitors). 

This control of the production of differentiated red blood cells by the macrophage at the level of the erythroblastic island may first appear in contradiction with usual knowledge of erythropoiesis. It is indeed well known that during a response to a stress, erythropoiesis is a very intense process, large amounts of cells being produced in a short time, so the whole process is expected to possess the ability to overcome its usual production, in particular the ratio of reticulocytes over progenitors should not stay constant. The hypothesis we propose is that the macrophage, inside the erythroblastic island, controls the explosiveness of erythropoiesis and that, contrary to other hematopoietic lineages (white cells, platelets), this control is necessary because stress erythropoiesis must deal with very large amounts of cells and otherwise the stability of the process could be lost. It is moreover noticeable that getting more stability with the central macrophage does not decrease responsiveness of the model. For instance, the island reacts to perturbations simulating anemia (sudden loss of mature red blood cells), returning quickly to `normal' values of the different erythroid populations, progenitors, reticulocytes and erythrocytes (not shown here). Hence the gain of stability is not compensated by a loss in the system's reactivity. Simulation of stress erythropoiesis (anemia) is actually a work in progress which should confirm that the macrophage plays a relevant role in the stability of erythropoiesis.

Some comments may however be made on the behavior of an island without macrophage in its center. Although it is clear that this system cannot lead to a stable island (this can be explained by considering that fluctuations automatically lead to proliferation or extinction), it is nevertheless possible to imagine a system working with such islands. For instance, if islands are put side-by-side, proliferation is made difficult (islands cannot expand due to confinement) and extinction can be compensated by the birth of new islands. Moreover, other assumptions, based on a better knowledge of erythroblastic islands, could lead to more stability (geometry of bone marrow and islands,\dots), even though all scenarii cannot and were not tested. However, the lack of stability makes it hard to have a clear view on how feedback controls act on one single island, and as far as we know of, studying erythropoiesis models (or hematopoiesis models, in a wide sense) without considering feedback controls does not make sense. Hence, searching for island stability appeared relevant in this study. At a global scale, it is still possible that island stability is not necessary and can be compensated between proliferation and extinction. When looking at one single island however, the fact that feedback controls do not work correctly on an unstable island is somewhat compelling.

Stability of the erythroblastic island including a central macrophage was obtained by considering two scenarii (see Figure \ref{fig-two-scenarii}) for the distribution of Erk and Fas quantities in daughter cells, following mitosis and mother cell's division. In the first one, initial values of Erk and Fas in newborn cells were set to zero, and cell fate was decided depending on the evolution of concentrations through the cell cycle duration (this is the `out of equilibrium' assumption). This scenario was shown to lead to results strongly dependent on the length of cell cycle. The second scenario, called the `hysteresis cycle case', which is more biologically relevant, consisted in dividing values of Erk and Fas of the mother cell at mitosis into two, so both daughter cells are located on a hysteresis cycle in the Erk-Fas plane at birth. Combining the hysteresis cycle scenario with a structure including a local feedback on the island (the macrophage) gave relevant results and appears very promising. Parameter values which were obtained through mathematical analysis appear to be very robust and the system exhibits qualitatively the same behavior for a wide range of parameters. The stability of the island allows a better control of its size and its subpopulations and, as a result, gives much more satisfying results on feedback tests.

Regarding parameter values used to study the model and check different scenarii (with or without macrophage in the center of the island, distribution of initial quantities of Erk and Fas), one must note that most of these values are mainly not accessible through experiments. Some others, such as for instance the time of survival of RBC, are however rather quite arbitrary and not necessarily in agreement with usual knowledge on erythropoiesis. Yet, results show that the behaviour may be satisfying from a biological point of view. Since a wide range of parameters is allowed, it is  expected that this system can be tuned to fit experimental data and numerically reproduce certain stress behavior, which will be the next step for modeling erythropoiesis.

Feedback functions considered in this study may also be questioned. Indeed, both erythropoietin's influence on erythroid progenitor self-renewal capacity and on apoptosis protection were taken as linear functions of the number of circulating red blood cells. Saturation effects should be envisaged when confronting the model to strong perturbations (for instance, a severe anemia), in order to be closer to the reality where responses of the organism to a loss of red blood cells have shown to be nonlinear \cite{cpgmg2008, cdgv2010}. In order to perform tests presented in the current study, in which the focus was on normal erythropoiesis, linear dependencies of the feedback controls on the red blood cell population were nevertheless probably not limiting. The question of the relevance of the protein regulatory network considered here may also be asked. Erythroid progenitor fate does not depend only on two proteins, Erk and Fas. However, considering a very `complete' regulatory network, with several proteins, does not appear appropriate, at least in order to perform first tests on the hybrid model. Complexity that would arise from such networks would certainly hide roles and influences of other actors, such as growth factors, cell-cell interaction, feedback controls, etc. The regulatory network could however be improved by considering some key proteins or receptors, such as GATA-1, involved in erythroid progenitor differentiation, or EpoR, the erythropoietin receptor, which plays a crucial role for erythroid progenitor proliferation and inhibition of apoptosis \cite{aispuru2008, socolovsky2007}.

One last point that has not been developed in this study deals with the geometry of the bone marrow. First, particular geometries, with obstacles or describing the porous aspect of the bone marrow, have not been included in this study. The hybrid model can deal with various geometries, however the choice was made to focus on the behavior of one single erythroblastic island and on feedback controls at global and local scales. Second, the hybrid model was numerically simulated in two dimensions. The case of simulations in three dimensions has not been developed. Qualitatively, such a change is not expected to modify the results, it should nevertheless be tested.

Although several aspects of erythropoiesis have not been considered in this model, we believe the analysis of feedback control role and relevance as well as the investigation of the role of a central macrophage in the center of an erythroblastic island improve previous models of erythropoiesis and give some insight in the way erythroid cell proliferation and differentiation can be controlled, in vitro and in vivo. Further developments are however necessary, and confronting the hybrid model with experimental data on stress situations is a next step to validate this new and certainly promising approach.

\section*{Aknowledgements}

Authors thank Prof. Mark Koury and all members of the INRIA Team DRACULA \\(http://dracula.univ-lyon1.fr) for fruitful discussions.

\appendix

\section{Parameter Estimation}\label{parameters}

We discuss here the parameter values used when investigating stability of the hybrid model in Section \ref{s-results}. Some parameters are clearly related to the discrete macroscopic model while some others are related to intracellular dynamics.

Choosing individual-based modelling creates many implicit parameters. Most of them can be chosen quite easily (cell radius, parameters of motion, etc.) and are not essential in the results of simulations. However, other parameters clearly change the overall cell behavior and have to be set carefully.

Let first consider parameters associated with extracellular dynamics (mainly growth factor release and action). Exact quantities of Fas-ligand, Epo and growth factors released by the macrophage are not fundamental, because their effect will be normalized at the intracellular scale (intracellular parameters controlling the sensitivity to these molecules). More important are the diffusion coefficients, for both Fas-ligand and growth factors released by the macrophage. These parameters are chosen in order to describe a short range action of Fas-ligand, and to locate growth factors around the macrophage. It is important that the range of diffusion is set according to the size of the cells in order to have a realistic short-diffusion for instance. The distribution of cell cycle lengths is also really important: it changes how the island expands, how cell cycles are locally correlated and plays a role in the fate of immature cells when we work out of equilibrium. Parameter values are shown in Table \ref{extparam}.

Intracellular parameters are chosen in order to optimize robustness of the system. It is noticeable that no experimental data is available to set values of these parameters, only the expected behavior is accessible. Parameters $b$ and $c$ of system (\ref{erk})--(\ref{fas}) control the relative and absolute speeds of the reactions. They are set first. The magnitude of the hysteresis cycle can then be tuned by choosing carefully $\alpha/b$ and $\beta/b$ (these ratios control its height, their absolute values its width). Parameter $a$ controls the offset of the cycle and is set such that it is located in the $[0,1] \times [0,1]$ domain. Parameter $d$ mainly controls how a cell goes from a stable branch of the cycle to the other: the higher the value of $d$, the higher the change in Fas quantity.

In equation (\ref{fas}), the function $\gamma$ explicitly describes the influence of Fas-ligand. This function has been taken as $\gamma = k_\gamma F_L$, where $k_\gamma$ is the sensitivity of progenitors to Fas-ligand.

Finally, $\alpha$ and $F_{cr}$ also depend on extracellular molecules. However, in what follows we will first consider the case when feedback loops are not included, so $\alpha$ and $F_{cr}$ will be constants first.

When Erk and Fas levels are supposed to be zero when the cell starts its cycle, the speeds of reactions are very important: some cells have to remain trapped near the origin at the end of their cycle in order to observe differentiation. The values of $b$ and $c$ are thus very constrained. The others simply define a large hysteresis cycle which aims at separating the different cell subpopulations as far as possible from each other. See Table \ref{intparam1} for values.

When Erk and Fas values are initially on the hysteresis cycle, then $b$ and $c$ are allowed to vary more largely. It has been decided to put quicker dynamics than in the first case in order to get rapid variations along the cycle. The width of the hysteresis cycle was also reduced, so values for $\alpha$ and $\beta$ are higher. See Table \ref{intparam2} for values.

Finally, an initial size for the island has to be set and the number of islands that will be simulated must be chosen. An initial size of 144 cells with central macrophage or 162 cells without macrophage was chosen, and parameters were adjusted to keep the size of the island roughly constant. Therefore, we only simulated one island at the beginning, but once reasonable parameters are found, it is possible to put several islands side-by-side and see how they interact. Preliminary experiments have been performed in this direction (not shown here), see the end of Section \ref{ss-stability}.


\newpage

\newpage

\newpage

\newpage

\newpage

\newpage

\newpage

\newpage

\newpage

\newpage

\newpage

\begin{table}[!ht]
\begin{center}
	\begin{tabular}{|c|c|c|}
		\hline  Parameter & Value & Unit \\
		\hline
		Cell cycle length & 24 & $h$ \\
		Cell cycle variations & 4/3 & $h$ \\
		$r_0$ & 0.01 & $L$ \\
		$m$ & 1 & $M$ \\
		$\mu$ & $3.10^5$ & $h^{-1}$ \\
		$K$ & $9.10^5$ & $M.h^{-2}$ \\
		\hline
		$D_{F_L}$ & $3.10^{-4}$ & $L^2.h^{-1}$ \\
		$k_{F_L}$ & $3.10^{-3}$ & molecules.cell$^{-1}$.$h^{-1}$ \\
		$\sigma_{F_L}$ & 0.6 & $h^{-1}$ \\
		\hline
		$D_{GF}$ & $3.10^{-3}$ & $L^2.h^{-1}$ \\
		$W_{GF}$ & $3.10^{-2}$ & molecules.$L^{-2}$.$h^{-1}$ \\
		$\sigma_{GF}$ & 0.3 & $h^{-1}$ \\
		\hline
	\end{tabular}
\end{center}
\caption{\textbf{Extracellular parameters} ($M$ is an arbitrary mass unit, $L$ an arbitrary length unit)} \label{extparam}
\end{table}

\newpage

\begin{table}[!ht]
\begin{center}
	\begin{tabular}{|c|c|c|}
		\hline  Parameter & Value & Unit \\
		\hline
		$\alpha$ & 0.01 & $h^{-1}$ \\
		$\beta$ & 1.5 & $h^{-1}.NU^{-2}$ \\
		$k$ & 2 & -\\
		$a$ & 0.12 & $h^{-1}$ \\
		$b$ & 0.414 & $h^{-1}.NU^{-1}$ \\
		\hline
		$k_\gamma$ & 0.009 & $h^{-1}.NU^{-1}$.molecule$^{-1}$ \\
		$c$ & 0.0828 & $h^{-1}.NU^{-1}$ \\
		$d$ & 0.006 & $h^{-1}$ \\
		\hline
		$E_{cr}$ & 0.31 & $NU$ \\
		$F_{cr}$ & 0.3 & $NU$ \\
		\hline
	\end{tabular}
\end{center}
\caption{\textbf{Internal parameters when cells begin at the origin (`out of equilibrium' case).} $NU$ is a normalized quantity unit for the intracellular molecules (maximum possible quantity is 1).} \label{intparam1}
\end{table}

\newpage

\begin{table}
\begin{center}
	\begin{tabular}{|c|c|c|}
		\hline  Parameter & Value & Unit \\
		\hline
		$\alpha$ & 1.62 & $h^{-1}$ \\
		$\beta$ & 60 & $h^{-1}.NU^{-1}$ \\
		$k$ & 2 & -\\
		$a$ & 15.9 & $h^{-1}$ \\
		$b$ & 1.8 & $h^{-1}.NU^{-1}$ \\
		\hline
		$k_\gamma$ & 0.9 & $h^{-1}.NU^{-2}$.molecule$^{-1}$ \\
		$c$ & 3 & $h^{-1}.NU^{-1}$ \\
		$d$ & 1.5 & $h^{-1}$ \\
		\hline
		$E_{cr}$ & 0.3 & $NU$ \\
		$F_{cr}$ & 0.6 & $NU$ \\
		\hline
	\end{tabular}
\end{center}
\caption{\textbf{Internal parameters when cells are on the hysteresis cycle.} $NU$ is a normalized quantity unit for the intracellular molecules (maximum possible quantity is 1).}\label{intparam2}
\end{table}

\end{document}